\documentclass[preprint,aps,11pt,showpacs,nofootinbib,tightenlines]{revtex4}
\usepackage{amsmath}
\usepackage{amssymb}
\usepackage{epsfig}
\usepackage{graphicx}
\begin{document}

\newcommand{\beq}{\begin{eqnarray}}
\newcommand{\eeq}{\end{eqnarray}}
\newcommand{\non}{\nonumber\\ }

\newcommand{\ov}{\overline}

\newcommand{\psl}{ P \hspace{-2.8truemm}/ }
\newcommand{\nsl}{ n \hspace{-2.2truemm}/ }
\newcommand{\vsl}{ v \hspace{-2.2truemm}/ }
\newcommand{\epsl}{\epsilon \hspace{-1.8truemm}/\,  }

\def \epjc{ Eur. Phys. J. C }
\def \jpg{  J. Phys. G }
\def \npb{  Nucl. Phys. B }
\def \plb{  Phys. Lett. B }
\def \pr{  Phys. Rep. }
\def \prd{  Phys. Rev. D }
\def \prl{  Phys. Rev. Lett.  }
\def \zpc{  Z. Phys. C  }
\def \jhep{ J. High Energy Phys.  }
\def \ijmpa { Int. J. Mod. Phys. A }

\title{Charmless hadronic $B_c \to VA, AA$ decays in the perturbative QCD approach}
\author{Xin Liu\footnote{liuxin.physics@gmail.com}
          and Zhen-Jun Xiao\footnote{xiaozhenjun@njnu.edu.cn}}
\affiliation{ Department of Physics and Institute of Theoretical
Physics, Nanjing Normal University, Nanjing, Jiangsu 210046,
People's Republic of China }
\date{\today}
\begin{abstract}
In this work, we calculate the branching ratios (BRs) and the
polarization fractions of sixty two charmless two-body $B_c$ meson
decays into final states involving one vector and one axial-vector
meson ($VA$) or two axial-vector mesons($AA$) within the framework
of perturbative QCD approach systematically, where $A$ is either a
$^3P_1$ or $^1P_1$ axial-vector meson. All considered decay
channels can only occur through the annihilation topologies in the
standard model. Based on the perturbative calculations and
phenomenological analysis, we find the following results: (i) the
{\it CP}-averaged BRs of the considered sixty two $B_c$ decays are
in the range of $10^{-5}$ to $10^{-9}$; 
(ii) since the behavior for $^1P_1$ meson is much different from
that of $^3P_1$ meson, the BRs of $B_c \to A(^1P_1) (V, A(^1P_1))$
decays are generally larger than that of $B_c \to A(^3P_1) (V,
A(^3P_1))$ decays in the perturbative QCD approach; 
(iii) many considered decays modes,
such as $B_c\to a_1(1260)^+ \omega$, $b_1(1235) \rho$, etc,
have sizable BRs within the reach of the LHCb experiments;
(iv) the longitudinal polarization fractions of most considered decays
are large and play the dominant role;
(v) the perturbative QCD predictions for several decays involving
mixtures of $^3P_1$ and/or $^1P_1$ mesons are highly sensitive to
the values of the mixing angles, which will be tested by the
ongoing LHC and forthcoming Super-B experiments; 
(vi) the {\it CP}-violating asymmetries of these considered $B_c$ decays
are absent in the standard model because only one type tree operator
is involved.
\end{abstract}


\maketitle

\section{Introduction}

Unlike the $B_q$ meson with $q=(u,d,s)$,  the $B_c$ meson is the
only heavy meson embracing two heavy quarks $b$ and $c$
simultaneously. Researchers believe that the $B_c$ physics must be
very rich if the statistics reaches high level. With the running
of Large Hadron
Collider(LHC) experiments, a great number of $B_c$ meson events, about
$10\%$ of the total $B$ meson data,  will be collected and this will
provide a new platform for both theorists and experimentalists to
study the perturbative and nonperturbative QCD dynamics, final
state interactions, even the new physics scenarios
beyond the standard model(SM) ~\cite{nb04:bcre}.

Very recently, we studied the two-body charmless hadronic decays $B_c \to PP, PV/VP, VV$ and
$B_c \to AP$ (here $P, V$ and $A$ stand for the light pseudo-scalar, vector and axial-vector
mesons respectively)~\cite{xiao09:bcnd,xiao10:bcap}. All these decays can only occur
via the annihilation type diagrams in the SM.
Although the contributions induced by annihilation diagrams are suppressed
in the decays of ordinary light $B_q$ ($q=u,d,s$) mesons, they
could be large and detected at LHC experiments~\cite{ekou09:ncbc} in $B_c$ meson decays.
According to the discussions as given in Ref.~\cite{ekou09:ncbc},
the charmless hadronic $B_c$ decays with decay rates at the level of
$10^{-6}$ could be measured at the LHC experiments with the accuracy required for the
phenomenological analysis, it is therefore believed that they
can help the people to understand the annihilation decay mechanism in
$B$ physics well. In this paper, we will extend our previous studies of two-body
$B_c$ decays to $B_c \to VA$ and $AA$ modes, which are also pure annihilation type decays, and
expected to have rich physics since there are three polarization
states involved in these decays.

In this paper, we will calculate the {\it CP}-averaged branching ratios (BRs)
and polarization fractions of the sixty two charmless hadronic $B_c \to VA,
AA$ decays by employing the low energy effective
Hamiltonian~\cite{Buras96:weak} and the perturbative QCD (pQCD) factorization approach
~\cite{Keum01:kpi,Lu01:pipi,Li03:ppnp}.
In the pQCD approach,  the annihilation type
diagrams can be calculated analytically, as have been done for example in
Refs.~\cite{Keum01:kpi,Lu01:pipi,Lu03:dsk,Li05:kphi,xiao06,Ali07:bsnd,xiao08:keta,Chen03:three}.
First of all, the size of annihilation contributions is an important
issue in the $B$ meson physics, and has been studied extensively, for example,
in Refs.~\cite{Lu03:dsk,Keum01:kpi,Lu01:pipi,Hong06:direct,Li05:kphi,Gritsan07:kphi}.
Second, the internal structure of the axial-vector mesons has
been one of the hot topics in recent
years~\cite{Lipkin77,Yang05,Yang07:twist}. Although many efforts
on both theoretical and experimental sides have been
made~\cite{Cheng07:ap,Yang07:ba1,Wang08:a1,Cheng08:aa,Li09:afm,Amsler08:pdg,Barberio08:hfag}
to explore it through the  studies for the relevant decay rates,
{\it CP}-violating asymmetries, polarization fractions and form
factors, etc., we currently still know little about the nature of
the axial-vector mesons. Furthermore, through the polarization
studies in the considered $B_c \to VA, AA$ decays, these channels
can shed light on the underlying helicity structure of the decay mechanism.

The paper is organized as follows. In Sec.~\ref{sec:1}, we present
the formalism and give the essential input quantities, including
the operator basis and the mixing angles between $^3P_1$ and/or $^1P_1$
mesons. The wave functions and distribution
amplitudes for $B_c$ and light vector and axial-vector
mesons are also given here. Then we perform the perturbative
calculations for considered decay channels in Sec.~\ref{sec:2}.
The analytic expressions of the decay amplitudes for all considered
sixty two $B_c \to VA, AA$ decay modes are also collected in this section. The
numerical results and phenomenological analysis are given in
Sec.~\ref{sec:3}. The main conclusions and a short summary are
presented in the last section.

\section{Input Quantities and Formalism} \label{sec:1}

In the following we shall briefly discuss the mixing of axial-vector mesons and summarize
all the input quantities relevant to the present work, such as the operator basis,
mixing angles, wave functions and light-cone distribution amplitudes for light
vector and axial-vector mesons. Finally, the formalism of pQCD approach
will also be presented briefly.

\subsection{Effective Hamiltonian}

For those considered charmless  hadronic $B_c$ decays, the related weak effective
Hamiltonian $H_{\rm eff}$ is given by~\cite{Buras96:weak}
\beq
\label{eq:heff}
H_{\rm eff} = \frac{G_{F}} {\sqrt{2}} \, \left[
V_{cb}^* V_{uD} \left (C_1(\mu) O_1(\mu) + C_2(\mu) O_2(\mu)
\right) \right] \;\label{heff} ,
\eeq
with the local four-quark tree operators $O_{1,2}$
\beq
O_1 &=& \bar u_\beta \gamma^\mu (1-\gamma_5)
D_\alpha \bar c_\beta \gamma^\mu (1- \gamma_5) b_\alpha \; , \non
O_2 &=& \bar u_\beta \gamma^\mu (1- \gamma_5) D_\beta \bar
c_\alpha \gamma^\mu (1- \gamma_5) b_\alpha \; ,
\eeq
where $V_{cb}, V_{uD}$ are the Cabibbo-Kobayashi-Maskawa (CKM) matrix elements,
"$D$" denotes the
light down quark $d$ or $s$ and $C_i(\mu)(i=1,2)$ are Wilson coefficients
at the renormalization scale $\mu$. For the Wilson coefficients
$C_{i}(\mu)$, we will also use the leading order
expressions, although the next-to-leading order calculations
already exist in the literature~\cite{Buras96:weak}. This is the
consistent way to cancel the explicit $\mu$ dependence in the
theoretical formulas. For the renormalization group evolution of
the Wilson coefficients from higher scale to lower scale, we use
the formulas as given in Ref.~\cite{Lu01:pipi} directly.

\subsection{Mixtures and Mixing Angles}

In the quark model, there exist two distinct types of light parity-even axial-vector
mesons, namely, $^3P_1$ ( $J^{\rm PC}=1^{++}$)
and $^1P_1$($J^{\rm PC}=1^{+-}$).
The $^3P_1$ nonet consists of $a_1(1260)$, $f_1(1285)$, $f_1(1420)$
and $K_{1A}$ states, while the $^1P_1$ nonet has $b_1(1235)$,
$h_1(1170)$, $h_1(1380)$ and $K_{1B}$ states.
In the SU(3) flavor limit, these mesons can not mix with each other.
Because the $s$ quark is
heavier than $u,d$ quarks, the physical mass eigenstates
$K_1(1270)$ and $K_1(1400)$ are not purely $1^3P_1$ or $1^1P_1$
states, but believed to be mixtures of $K_{1A}$
and $K_{1B}$\footnote{For the sake of simplicity, we will
adopt the forms $a_1$, $b_1$, $K'$, $K^{''}$, $f'$, $f^{''}$, $h'$
and $h^{''}$ to denote the axial-vector mesons $a_1(1260)$,
$b_1(1235)$, $K_1(1270)$, $K_1(1400)$, $f_1(1285)$, $f_1(1420)$,
$h_1(1170)$ and $h_1(1380)$ correspondingly in the following
sections, unless otherwise stated. We will also use
$K_1$, $f_1$ and $h_1$ to denote $K_1(1270)$ and $K_1(1400)$, $f_1(1285)$
and $f_1(1420)$, and $h_1(1170)$ and $h_1(1380)$ for convenience
unless explicitly otherwise stated.}.
Analogous to $\eta$ and
$\eta^\prime$ system, the flavor-singlet and flavor-octet axial-vector
meson can also mix with each other. It is worth mentioning that
the mixing angles can be determined by the relevant data, but
unfortunately, there is no enough data now for these mesons which
leaves the mixing angles basically free parameters.

The physical states $K_1(1270)$ and $K_1(1400)$ can be written as the mixtures of the
$K_{1A}$ and $K_{1B}$ states:
\beq
\left ( \begin{array}{l} K_1(1270)\\ K_1(1400)\\ \end{array} \right ) =
\left ( \begin{array}{rr}
\sin\theta_K & \cos\theta_K\\ \cos\theta_K& -\sin\theta_K\\ \end{array} \right )
\left ( \begin{array}{l} K_{1A}\\ K_{1B}\\ \end{array} \right )
\label{eq:k1mixing}
\eeq
The mixing angle $\theta_K$ still not be well determined because of the poor
experimental data. In this paper, for simplicity, we will adopt two reference
values as those used in Ref.~\cite{Yang07:twist}: $\theta_K=\pm 45^\circ$.

Analogous to the $\eta$-$\eta'$ mixing in the pseudoscalar sector,
the $h_1(1170)$ and $h_1(1380)$ ($1^1P_1$ states) system
can be mixed in terms of the pure singlet $|h_1\rangle$ and octet $|h_8\rangle$,
\beq
\left ( \begin{array}{l} h_1(1170)\\ h_1(1380)\\ \end{array} \right ) =
\left ( \begin{array}{rr}
\cos\theta_1 & \sin\theta_1\\ -\sin\theta_1& \cos\theta_1\\ \end{array} \right )
\left ( \begin{array}{l} h_1\\ h_8\\ \end{array} \right )
\label{eq:h1mixing}
\eeq
Likewise, $f_1(1285)$ and $f_1(1420)$ (the $1^3P_1$ states) will mix in the form of
\beq
\left ( \begin{array}{l} f_1(1285)\\ f_1(1420)\\ \end{array} \right ) =
\left ( \begin{array}{rr}
\cos\theta_3 & \sin\theta_3\\ -\sin\theta_3& \cos\theta_3\\ \end{array} \right )
\left ( \begin{array}{l} f_1\\ f_8\\ \end{array} \right )
\label{eq:f1mixing}
\eeq
where the component of $h_1, f_1$ and $h_8, f_8$
can be written as
\beq
 |h_1\rangle = |f_1\rangle &=&\frac{1}{\sqrt{3}} \left ( |\bar q q \rangle
+ |\bar ss\rangle\right ), \non
 |h_8\rangle= |f_8\rangle &=& \frac{1}{\sqrt{6}}\left (|\bar qq\rangle
-2 |\bar ss\rangle \right ),
\eeq
where $q=(u,d)$. The values of the mixing angles for $1^1P_1$ and $1^3P_1$ states
are chosen as \cite{Yang07:twist}:
\beq
\theta_1=10^\circ \quad or \quad 45^\circ; \qquad
\theta_3=38^\circ \quad or \quad 50^\circ.
\eeq

\subsection{Wave Functions and Distribution Amplitudes}\label{ssec:wf}

In order to calculate the decay amplitude, we should choose the
proper wave functions for the heavy $B_c$, and light vector and axial-vector mesons.
For the wave function of $B_c$ meson, we
adopt the form(see Ref.~\cite{xiao09:bcnd}, and references therein) as,
\beq
\Phi_{B_c} (x) &=& \frac{i}{\sqrt{2 N_c}}\left[ (\psl  + m_{B_c})
\gamma_5 \phi_{B_c}(x) \right]_{\alpha\beta}\;.
\eeq
where the distribution amplitude $\phi_{B_c}$ would be close to
$\delta(x-m_c/m_{B_c})$ in the non-relativistic limit because of
the fact that $B_c$ meson embraces two heavy quarks. 
We therefore adopt the non-relativistic approximation form for $\phi_{B_c}$
as~\cite{CDL,SYDM}, \beq \phi_{B_c}(x) &=& \frac{f_{B_c}}{2
\sqrt{2 N_c}} \delta (x- m_c/m_{B_c})\;, \eeq where $f_{B_c}$ and
$N_c$ are the decay constant of $B_c$ meson and the color number,
respectively.

For the wave functions of vector and axial-vector mesons, one
longitudinal($L$) and two transverse($T$) polarizations are
involved, and can be written as,
\beq
\Phi^L_V(x)&=&
\frac{1}{\sqrt{2 N_c}} \left\{ m_V \epsl_V^{*L} \phi_V(x) +
\epsl^{*L}_V \psl \phi_V^t(x)+ m_V
\phi_V^s(x)\right\}_{\alpha\beta} \;, \\
\Phi^T_V(x)&=&\frac{1}{\sqrt{2 N_c}} \left\{ m_V \epsl_V^{*T} \phi_V^v(x) +
\epsl^{*T}_V \psl \phi_V^T(x)+ m_V i \epsilon_{\mu\nu\rho\sigma}
\gamma_5 \gamma^\mu \epsilon_T^{*\nu} n^\rho v^\sigma
\phi_V^a(x)\right\}_{\alpha\beta} \;,
\eeq
\beq
\Phi^L_A(x)&=& \frac{1}{\sqrt{2 N_c}}\gamma_5 \left\{
m_A \epsl_A^{*L} \phi_A(x) + \epsl^{*L}_A \psl \phi_A^t(x)+ m_A
\phi_A^s(x)\right\}_{\alpha\beta} \;, \\
\Phi^T_A(x)&=&\frac{1}{\sqrt{2 N_c}} \gamma_5\left\{ m_A
\epsl_A^{*T} \phi_A^v(x) + \epsl^{*T}_A \psl \phi_A^T(x)+ m_A i
\epsilon_{\mu\nu\rho\sigma} \gamma_5 \gamma^\mu \epsilon_T^{*\nu}
n^\rho v^\sigma \phi_A^a(x)\right\}_{\alpha\beta} \;,
\eeq
where $\epsilon_{V(A)}^{L,T}$ denotes the longitudinal and
transverse polarization vectors of vector(axial-vector) meson,
satisfying $P \cdot \epsilon=0$ in each polarization, $x$ denotes
the momentum fraction carried by quark in the meson, and
$n=(1,0,{\bf 0}_T)$ and $v=(0,1,{\bf 0}_T)$ are dimensionless
light-like unit vectors. We here adopt the convention
$\epsilon^{0123}=1$ for the Levi-Civita tensor
$\epsilon^{\mu\nu\alpha\beta}$.

The twist-2 distribution amplitudes for the longitudinally and
tranversely polarized vector meson can be parameterized as:
\beq
\phi_{V}(x)&=&\frac{3f_{V}}{\sqrt{2N_c}} x
(1-x)\left[1+3a_{1V}^{||}\, (2x-1)+ a_{2V}^{||}\, \frac{3}{2} ( 5(2x-1)^2  - 1 )\right]\;,\label{eq:ldav}\\
\phi_{V}^T(x)&=&\frac{3f^T_{V}}{\sqrt{2N_c}} x
(1-x)\left[1+3a_{1V}^{\perp}\, (2x-1)+ a_{2V}^{\perp}\,
\frac{3}{2} ( 5(2x-1)^2  - 1 )\right]\;,\label{eq:tdav}
\eeq
Here $f_{V}$ and $f_V^T$ are the decay constants of the vector meson
with longitudinal and tranverse polarization, respectively.

The Gegenbauer moments have been studied extensively in the
literatures \cite{rho,previousvectorwf}, here we adopt the following
values from the recent updates~\cite{adoptedvectorwf,LCSRBZ,twist3}:
\beq
a_{1K^*}^{||}&=&0.03\pm0.02,a_{2K^*}^{||}=0.11\pm0.09,
a_{2\rho}^{||}=a_{2\omega}^{||}=0.15\pm0.07,
a_{2\phi}^{||}=0.18\pm0.08\;;\\
a_{1K^*}^\perp&=&0.04\pm0.03,a_{2K^*}^{\perp}=0.10\pm0.08,
a_{2\rho}^{\perp}=a_{2\omega}^{\perp}=0.14\pm0.06,
a_{2\phi}^{\perp}=0.14\pm0.07\;.
\eeq

The asymptotic forms of the twist-3 distribution amplitudes
$\phi^{t,s}_V$ and $\phi_V^{v,a}$ are \cite{Li05:kphi}: \beq
\phi^t_V(x) &=& \frac{3f^T_V}{2\sqrt {2N_c}}(2x-1)^2,\;\;\;\;\;\;\;\;\;\;\;
  \hspace*{0.5cm} \phi^s_V(x)=-\frac{3f_V^T}{2\sqrt {2N_c}} (2x-1)~,\\
\phi_V^v(x)&=&\frac{3f_V}{8\sqrt{2N_c}}(1+(2x-1)^2),\;\;\; \ \ \
 \phi_V^a(x)=-\frac{3f_V}{4\sqrt{2N_c}}(2x-1).
\eeq

The twist-2 distribution amplitudes for the longitudinally and trasversely
polarized axial-vector $^3P_1$ and $^1P_1$ mesons can be
parameterized as~\cite{Yang07:twist,Li09:afm}:
\beq
 \phi_A(x) & = & \frac{3 f}{ \sqrt{2 N_c}}  x (1- x) \left[ a_{0A}^\parallel + 3
a_{1A}^\parallel\, (2x-1) +
a_{2A}^\parallel\, \frac{3}{2} ( 5(2x-1)^2  - 1 ) \right] ,\label{eq:ldaa}\\
 \phi_A^T(x) & = & \frac{3 f}{ \sqrt{2 N_c}}  x (1- x) \left[ a_{0A}^\perp + 3 a_{1A}^\perp\, (2x-1) +
a_{2A}^\perp\, \frac{3}{2} ( 5(2x-1)^2  - 1 ) \right], \label{eq:tdaa}
\eeq
Here, the definition of these distribution amplitudes $\phi_A(x)$ and $\phi_A^T(x)$ satisfy the
 following relations:
 \beq
\int_0^1 \phi_{^3P_1}(x) &=& \frac{f_{^3P_1}}{2 \sqrt{2 N_c}}, \;\;\;\;\;\;\;\;\;\;\;\;\;\;\;\;
\int_0^1 \phi^T_{^3P_1}(x) = a^{\perp}_{0 ^3P_1}\frac{f_{^3P_1}}{2 \sqrt{2 N_c}}\;;\non
\int_0^1 \phi_{^1P_1}(x) &=& a^{||}_{0 ^1P_1} \frac{f_{^1P_1}}{2 \sqrt{2 N_c}}, \;\;\;\;\;\;\;\;
\int_0^1 \phi^T_{^1P_1}(x) =  \frac{f_{^1P_1}}{2 \sqrt{2 N_c}}\;.
 \eeq
 where $a^{||}_{0 ^3P_1}= 1$ and $a^{\perp}_{0 ^1P_1}= 1$ have been used.

As for twist-3 distribution amplitudes for axial-vector meson, we use the following form~\cite{Li09:afm}:
\beq
\phi_{A}^t(x) &= &\frac{3 f}{2\sqrt{2N_c}}\left\{ a_{0A}^\perp (2x-1)^2+ \frac{1}{2}\,a_{1A}^\perp\,(2x-1) (3 (2x-1)^2-1) \right\}
 ,\\
\phi_{A}^s(x)&=& \frac{3 f}{2\sqrt{2N_c}} \frac{d}{dx}\left\{ x (1- x) ( a_{0A}^\perp + a_{1A}^\perp (2x-1) ) \right\}.
 \eeq
 \beq
 \phi_{A}^v(x)&=&\frac{3 f}{4\sqrt{2N_c}} \left\{ \frac{1}{2} a_{0A}^\parallel (1+(2x-1)^2) +  a_{1A}^\parallel (2x-1)^3 \right\}
 , \\
 \phi_{A}^a(x)&=& \frac{3 f}{4\sqrt{2N_c}}\frac{d}{dx}  \left\{ x (1- x) ( a_{0A}^\parallel + a_{1A}^\parallel (2x-1))  \right\}\;.
 \eeq
 where $f$ is the decay constant. 
 It should be noted that in the above distribution amplitudes of strange axial-vector
 mesons $K_{1A}$ and $K_{1B}$, $x$ stands for the momentum fraction carrying by the $s$ quark.

The Gegenbauer moments have been studied extensively in the
literatures (see Ref.~\cite{Yang07:twist} and references therein), here we adopt the following
values:
\beq
 a^{||}_{2a_1}&=& -0.02\pm 0.02;\;\;\;\;\;\;\;\;\;  a^{\perp}_{1a_1}= -1.04\pm 0.34;\;\;\;\;\;\;\;\;\;  a^{||}_{1b_1}=  -1.95\pm 0.35; \non
 a^{||}_{2f_1} &=& -0.04\pm 0.03;\;\;\;\;\;\;\;\;\; a^{\perp}_{1f_1}= -1.06\pm 0.36;\;\;\;\;\;\;\;\;\;  a^{||}_{1h_1}= -2.00\pm 0.35;\non
 a^{||}_{2f_8} &=& -0.07\pm 0.04;\;\;\;\;\;\;\;\;\; a^{\perp}_{1f_8}= -1.11\pm 0.31;\;\;\;\;\;\;\;\;\;  a^{||}_{1h_8}= -1.95\pm 0.35;\non
 a^{||}_{1K_{1A}}&=& 0.00\pm 0.26;\;\;\;\;\;\;\;\;\; a^{||}_{2K_{1A}}= -0.05\pm 0.03;\;\;\;\;\;\;\; a^{\perp}_{0K_{1A}}= 0.08\pm 0.09;\non
 a^{\perp}_{1K_{1A}}&=& -1.08\pm 0.48;\;\;\;\;\;\; a^{||}_{0K_{1B}}= 0.14\pm 0.15;\;\;\;\;\;\;\;\;\;\; a^{||}_{1K_{1B}}= -1.95\pm 0.45;\non
 a^{||}_{2K_{1B}}&=& 0.02\pm 0.10;\;\;\;\;\;\;\;\;\; a^{\perp}_{1K_{1B}}=0.17\pm 0.22.
\eeq

\subsection{Formalism of pQCD approach}

Since the $b$ quark is rather heavy, we work in the frame with the
$B_c$ meson at rest, i.e., with the $B_c$ meson momentum
$P_1=(m_{B_c}/\sqrt{2})(1,1,{\bf 0}_T)$ in the light-cone
coordinates. For the non-leptonic charmless $B_c \to M_2 M_3$\footnote{For the sake of simplicity, in the following, we will
use $M_2$ and $M_3$ to denote the final state mesons
respectively, unless otherwise stated. } decays, assuming that the
$M_2$ ($M_3$) meson moves in the plus (minus) $z$ direction
carrying the momentum $P_2$ ($P_3$) and the polarization vector
$\epsilon_2$ ($\epsilon_3$). Then the two final state meson
momenta can be written as
\beq
     P_2 =\frac{m_{B_c}}{\sqrt{2}} (1-r_3^2,r_2^2,{\bf 0}_T), \quad
     P_3 =\frac{m_{B_c}}{\sqrt{2}} (r_3^2,1-r_2^2,{\bf 0}_T),
\eeq
respectively, where $r_2=m_2/m_{B_c}$, $r_3=m_3/m_{B_c}$ with $m_2=m_{M_2}$ and $m_3=m_{M_3}$.
The longitudinal polarization vectors,
$\epsilon_2^L$ and $\epsilon_3^L$, can be given by
\beq
\epsilon_2^L =\frac{m_{B_c}}{\sqrt{2}m_2} (1-r_3^2,
-r_2^2,{\bf 0}_T), \quad \epsilon_3^L =
\frac{m_{B_c}}{\sqrt{2}m_3} (-r_3^2, 1-r_2^2,{\bf0}_T).
\eeq
And the transverse ones are parameterized as $\epsilon_2^T = (0,
0,1_T)$, and $\epsilon_3^T = (0, 0,1_T)$. Putting the (light)
 quark momenta in $B_c$, $M_2$ and $M_3$ mesons as $k_1$,
$k_2$, and $k_3$, respectively, we can choose
\beq
k_1 = (x_1P_1^+,0,{\bf k}_{1T}), \quad k_2 = (x_2 P_2^+,0,{\bf k}_{2T}),
\quad k_3 = (0, x_3 P_3^-,{\bf k}_{3T}).
\eeq
Then, for $B_c \to M_2 M_3$ decays, the integration over $k_1^-$, $k_2^-$, and
$k_3^+$ will conceptually lead to the decay amplitudes in the pQCD
approach,
\beq
{\cal A}(B_c \to M_2 M_3) &\sim &\int\!\! d x_1 d
x_2 d x_3 b_1 d b_1 b_2 d b_2 b_3 d b_3 \non && \cdot \mathrm{Tr}
\left [ C(t) \Phi_{B_c}(x_1,b_1) \Phi_{M_2}(x_2,b_2)
\Phi_{M_3}(x_3, b_3) H(x_i, b_i, t) S_t(x_i)\, e^{-S(t)} \right
]. \label{eq:a2}
\eeq
where $b_i$ is the conjugate space
coordinate of $k_{iT}$, and $t$ is the largest energy scale in
function $H(x_i,b_i,t)$. The large logarithms $\ln (m_W/t)$ are
included in the Wilson coefficients $C(t)$. The large double
logarithms ($\ln^2 x_i$) are summed by the threshold
resummation~\cite{Li02:resum}, and they lead to $S_t(x_i)$ which
smears the endpoint singularities on $x_i$. The last term,
$e^{-S(t)}$, is the Sudakov form factor which suppresses the soft
dynamics effectively~\cite{Li98:soft}. Thus it makes the
perturbative calculation of the hard part $H$ applicable at
intermediate scale, i.e., $m_{B_c}$ scale. We will calculate
analytically the function $H(x_i,b_i,t)$ for the considered decays
at leading order in $\alpha_s$ expansion and give the
convoluted amplitudes in next section.

\section{Perturbative Calculations in pQCD approach} \label{sec:2}

\begin{figure}[t,b]
\vspace{-0.5cm} \centerline{\epsfxsize=13 cm \epsffile{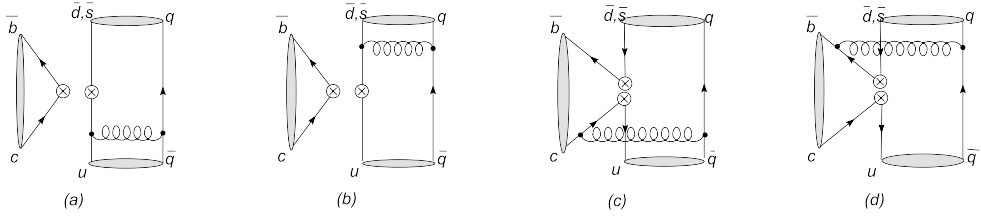}}
\vspace{0.2cm} \caption{Typical Feynman diagrams for charmless hadronic $B_c \to VA,AA$ decays.}
 \label{fig:fig1}
\end{figure}

There are three kinds of polarizations of a vector or axial-vector
meson, namely, longitudinal ($L$), normal ($N$), and transverse
($T$). Similar to the pure annihilation type $B_c \to VV$
decays~\cite{xiao09:bcnd}, the amplitudes for a $B_c$ meson
decaying into one vector and one axial-vector meson or two
axial-vector mesons are also characterized by the polarization
states of these vector and axial-vector mesons. In terms of helicities,
the decay amplitudes ${\cal M}^{(\sigma)}$ for $B_c
\to M_2(P_2,\epsilon^*_2) M_3(P_3,\epsilon^*_3)$ decays can be
generally described by
\beq
{\cal M}^{(\sigma)}&=&\epsilon_{2\mu}^{*}(\sigma)\epsilon_{3\nu}^{*}(\sigma) \left[ a
\,\, g^{\mu\nu} + {b \over m_2 m_3} P_1^\mu P_1^\nu + i{c
\over m_2 m_3 } \epsilon^{\mu\nu\alpha\beta} P_{2\alpha}
P_{3\beta}\right]\;,\non
&\equiv &m_{B_c}^{2}{\cal M}_{L}+m_{B_c}^{2}{\cal M}_{N}
\epsilon^{*}_{2}(\sigma=T)\cdot\epsilon^{*}_{3}(\sigma=T) \non &&
+i{\cal M}_{T}\epsilon^{\alpha \beta\gamma \rho}
\epsilon^{*}_{2\alpha}(\sigma)\epsilon^{*}_{3\beta}(\sigma)
P_{2\gamma }P_{3\rho }\; , \label{eq:msigma}
\eeq
where the superscript $\sigma$ denotes the helicity states of one vector and one axial-vector meson or two
axial-vector mesons with $L(T)$ standing for the longitudinal (transverse)
component. And the definitions of the amplitudes $ {\cal M}_{i}
(i=L,N,T)$ in terms of the Lorentz-invariant amplitudes $a$, $b$
and $c$ are
\beq
m_{B_c}^2 \,\, {\cal M}_L &=& a \,\,
\epsilon_2^{*}(L) \cdot \epsilon_3^{*}(L) +{b \over m_2
m_3 } \epsilon_{2}^{*}(L) \cdot P_3 \,\, \epsilon_{3}^{*}(L)
\cdot P_2\;, \non
m_{B_c}^2 \,\, {\cal M}_N &=& a \;,\label{eq:amp}\\
m_{B_c}^2 \,\, {\cal M}_T &=& {c \over r_2\, r_3}\;.\label{id-rel}\nonumber
\eeq
We therefore will evaluate the helicity amplitudes ${\cal M}_L,
{\cal M}_N, {\cal M}_T$ based on the pQCD factorization approach,
respectively.

In the following we will present analytically the factorization
formulas for sixty two charmless hadronic $B_c \to AV/VA,AA$ decays.
From the effective Hamiltonian~(\ref{heff}), there are four types of
diagrams contributing to these considered decays as illustrated in
Fig.~\ref{fig:fig1} with single $(V-A)(V-A)$ currents. From the
first two diagrams (a) and (b) in Fig.~\ref{fig:fig1}, by
perturbative QCD calculations, we can obtain the Feynman decay amplitudes
for factorizable annihilation contributions for $B_c \to AV$,
$VA$, $AA$ as the following sequence,
\beq
F^{L}_{fa}(AV) &=& -8 \pi C_F m_{B_c}^2 \int_0^1 d x_{2} dx_{3}\, \int_{0}^{\infty} b_2 db_2 b_3 db_3\,
\non && \times
\left\{ \left[x_{2} \phi_{A}(x_2)\phi_{V}(x_3) - 2 r_2^A r_3^V \left((x_2 + 1)\phi^{s}_{A}(x_2)+ (x_2 - 1)\phi^{t}_{A}(x_{2})\right)
\right.\right.\non && \left.\left. \times
\phi_{V}^s(x_3)\right]E_{fa}(t_{a}) h_{fa}(1-x_{3},x_{2},b_{3},b_{2})+E_{fa}(t_{b}) h_{fa}(x_{2},1-x_{3},b_{2},b_{3})
 \right. \non && \left. \times
\left[ (x_3 -1) \phi_{A}(x_2) \phi_{V}(x_3) - 2 r_2^A r_3^V \phi_{A}^s(x_2) \left( (x_3 -2)\phi_{V}^s(x_3)- x_3\phi_{V}^t(x_3)\right)\right] \right\}\;,
\label{eq:abl-av} \\
F^{N}_{fa}(AV) &=& -8 \pi C_F m_{B_c}^2 \int_0^1 d x_{2} dx_{3}\, \int_{0}^{\infty} b_2 db_2 b_3 db_3\,r_2^A r_3^V
\non && \times
\left\{ h_{fa}(1-x_{3},x_{2},b_{3},b_{2})E_{fa}(t_{a})\left[ (x_{2}+1 ) (\phi_{A}^a(x_2)\phi^a_{V}(x_3) + \phi_{A}^v(x_2)\phi^v_{V}(x_3))
\right.\right.\non && \left.\left.
+ (x_2 - 1) (\phi^{v}_{A}(x_{2})\phi_{V}^a(x_3)+\phi^{a}_{A}(x_{2})\phi_{V}^v(x_3))\right]
 \right. \non
 && \left.
+ \left[(x_3 -2) (\phi_{A}^a(x_2) \phi_{V}^a(x_3)+\phi_{A}^v(x_2) \phi_{V}^v(x_3) ) - x_3
\left(\phi_{A}^a(x_2)\phi_{V}^v(x_3)+\phi_{A}^v(x_2)\phi_{V}^a(x_3)\right)\right]
\right.
\non &&
\left. \hspace{0.3cm}\times E_{fa}(t_{b})h_{fa}(x_{2},1-x_{3},b_{2},b_{3})  \right\}\;
\label{eq:abn-av} \\
F^{T}_{fa}(AV) &=& -16 \pi C_F m_{B_c}^2 \int_0^1 d x_{2} dx_{3}\, \int_{0}^{\infty} b_2 db_2 b_3 db_3\,r_2^A r_3^V
\non && \times
\left\{ h_{fa}(1-x_{3},x_{2},b_{3},b_{2})E_{fa}(t_{a})\left[ (x_{2}+1 ) (\phi_{A}^a(x_2)\phi^v_{V}(x_3) + \phi_{A}^v(x_2)\phi^a_{V}(x_3))
\right.\right.\non && \left.\left.
+ (x_2 - 1)(\phi^{a}_{A}(x_{2})\phi_{V}^a(x_3)+\phi^{v}_{A}(x_{2})\phi_{V}^v(x_3))\right]
\right. \non
&& \left.
+ \left[(x_3 -2) (\phi_{A}^a(x_2) \phi_{V}^v(x_3)+\phi_{A}^v(x_2) \phi_{V}^a(x_3) ) - x_3
\left(\phi_{A}^a(x_2)\phi_{V}^a(x_3)+\phi_{A}^v(x_2)\phi_{V}^v(x_3)\right)\right]\right.\non
&& \left. \hspace{0.3cm}\times
E_{fa}(t_{b}) \; h_{fa}(x_{2},1-x_{3},b_{2},b_{3}) \right\}\;
\label{eq:abt-av}
\eeq
\beq
F^{L}_{fa}(VA) &=& -8 \pi C_F m_{B_c}^2 \int_0^1 d x_{2} dx_{3}\, \int_{0}^{\infty} b_2 db_2 b_3 db_3\,
\non && \times
\left\{ \left[x_{2} \phi_{V}(x_2)\phi_{A}(x_3) + 2 r_2^V r_3^A \left((x_2 + 1)\phi^{s}_{V}(x_2)+ (x_2 - 1)\phi^{t}_{V}(x_{2})\right)
\right.\right.\non && \left.\left.\times
\phi_{A}^s(x_3)\right]E_{fa}(t_{a}) h_{fa}(1-x_{3},x_{2},b_{3},b_{2})+E_{fa}(t_{b}) h_{fa}(x_{2},1-x_{3},b_{2},b_{3})
 \right. \non && \left. \times
\left[ (x_3 -1) \phi_{V}(x_2) \phi_{A}(x_3) + 2 r_2^V r_3^A \phi_{V}^s(x_2) \left( (x_3 -2)\phi_{A}^s(x_3)- x_3 \phi_{A}^t(x_3)\right)\right] \right\}\;,
\ \ \ \label{eq:abl-va}
\eeq
\beq
F^{N}_{fa}(VA) &=& -8 \pi C_F m_{B_c}^2 \int_0^1 d x_{2} dx_{3}\, \int_{0}^{\infty} b_2 db_2 b_3 db_3\,r_2^V r_3^A
\non && \times
\left\{ h_{fa}(1-x_{3},x_{2},b_{3},b_{2})E_{fa}(t_{a})\left[ (x_{2}+1 )(\phi_{V}^a(x_2)\phi^a_{A}(x_3) + \phi_{V}^v(x_2)\phi^v_{A}(x_3))
\right.\right.\non && \left.\left.
+ (x_2 - 1) (\phi^{v}_{V}(x_{2})\phi_{A}^a(x_3)+\phi^{a}_{V}(x_{2})\phi_{A}^v(x_3))\right]\right.\non
&& \left.
+  \left[(x_3 -2) (\phi_{V}^a(x_2) \phi_{A}^a(x_3)+\phi_{V}^v(x_2) \phi_{A}^v(x_3) ) - x_3 \left(\phi_{V}^a(x_2)\phi_{A}^v(x_3)
 +\phi_{V}^v(x_2)\phi_{A}^a(x_3)\right)\right] \right. \non
 && \left.
 \hspace{0.3cm}\times E_{fa}(t_{b})h_{fa}(x_{2},1-x_{3},b_{2},b_{3}) \right\}\;,
 \label{eq:abn-va} \\
F^{T}_{fa}(VA) &=& - 16 \pi C_F m_{B_c}^2 \int_0^1 d x_{2} dx_{3}\, \int_{0}^{\infty} b_2 db_2 b_3 db_3\,r_2^V r_3^A
\non && \times
\left\{ h_{fa}(1-x_{3},x_{2},b_{3},b_{2})E_{fa}(t_{a})\left[ (x_{2}+1 ) (\phi_{V}^a(x_2)\phi^v_{A}(x_3) + \phi_{V}^v(x_2)\phi^a_{A}(x_3))
\right.\right.\non && \left.\left.
+ (x_2 - 1)(\phi^{a}_{V}(x_{2})\phi_{A}^a(x_3)+\phi^{v}_{V}(x_{2})\phi_{A}^v(x_3))\right] \right. \non
&& \left.
+ \left[(x_3 -2) (\phi_{V}^a(x_2) \phi_{A}^v(x_3)+\phi_{V}^v(x_2) \phi_{A}^a(x_3) ) - x_3 \left(\phi_{V}^a(x_2)\phi_{A}^a(x_3)
+\phi_{V}^v(x_2)\phi_{A}^v(x_3)\right)\right] \right.\non
&& \left.
\hspace{0.3cm}\times E_{fa}(t_{b})\; h_{fa}(x_{2},1-x_{3},b_{2},b_{3}) \right\}\;,
\label{eq:abt-va}
\eeq
\beq
F^{L}_{fa}(AA) &=& 8 \pi C_F m_{B_c}^2 \int_0^1 d x_{2} dx_{3}\, \int_{0}^{\infty} b_2 db_2 b_3 db_3\,
\non && \times
\left\{ \left[x_{2} \phi_{2}(x_2)\phi_{3}(x_3) + 2 r_2^A r_3^A \left((x_2 + 1)\phi^{s}_{2}(x_2)+ (x_2
 -1)\phi^{t}_{2}(x_{2})\right)\right.\right.\non && \left.\left. \times \phi_{3}^s(x_3)\right]
E_{fa}(t_{a})  h_{fa}(1-x_{3},x_{2},b_{3},b_{2})+E_{fa}(t_{b})h_{fa}(x_{2},1-x_{3},b_{2},b_{3})
\right. \non && \left. \times
\left[ (x_3 -1) \phi_{2}(x_2) \phi_{3}(x_3) + 2 r_2^A r_3^A \phi_{2}^s(x_2) \left( (x_3 -2)\phi_{3}^s(x_3)
- x_3 \phi_{3}^t(x_3)\right)\right] \right\}\;, \label{eq:abl-aa}\\
F^{N}_{fa}(AA) &=& 8 \pi C_F m_{B_c}^2 \int_0^1 d x_{2} dx_{3}\, \int_{0}^{\infty} b_2 db_2 b_3 db_3\,r_2^A r_3^A
 \non && \times
\left\{h_{fa}(1-x_{3},x_{2},b_{3},b_{2})E_{fa}(t_{a})\left[ (x_{2}+1 ) (\phi_{2}^a(x_2)\phi^a_{3}(x_3) + \phi_{2}^v(x_2)\phi^v_{3}(x_3))
 \right.\right.\non && \left.\left.
+ (x_2 - 1) (\phi^{v}_{2}(x_{2})\phi_{3}^a(x_3)+\phi^{a}_{2}(x_{2})\phi_{3}^v(x_3))\right]
  \right. \non && \left.
  +   \left[(x_3 -2) (\phi_{2}^a(x_2) \phi_{3}^a(x_3)+\phi_{2}^v(x_2) \phi_{3}^v(x_3) ) - x_3
\left(\phi_{2}^a(x_2)\phi_{3}^v(x_3)+\phi_{2}^v(x_2)\phi_{3}^a(x_3)\right)\right]
\right. \non
&& \left.
\hspace{0.3cm}\times E_{fa}(t_{b})h_{fa}(x_{2},1-x_{3},b_{2},b_{3})\right\}\;,
\label{eq:abn-aa} \\
F^{T}_{fa}(AA) &=& 16 \pi C_F m_{B_c}^2 \int_0^1 d x_{2} dx_{3}\, \int_{0}^{\infty} b_2 db_2 b_3 db_3\,r_2^A r_3^A
\non && \times
\left\{ h_{fa}(1-x_{3},x_{2},b_{3},b_{2})E_{fa}(t_{a})\left[ (x_{2}+1 )
(\phi_{2}^a(x_2)\phi^v_{3}(x_3) + \phi_{2}^v(x_2)\phi^a_{3}(x_3))
\right.\right.\non && \left.\left.
+ (x_2 - 1)(\phi^{a}_{2}(x_{2})\phi_{3}^a(x_3)+\phi^{v}_{2}(x_{2})\phi_{3}^v(x_3))\right]
 \right. \non
 && \left.
+ \left[(x_3 -2) (\phi_{2}^a(x_2) \phi_{3}^v(x_3)+\phi_{2}^v(x_2) \phi_{3}^a(x_3) ) - x_3
 \left(\phi_{2}^a(x_2)\phi_{3}^a(x_3)+\phi_{2}^v(x_2)\phi_{3}^v(x_3)\right)\right]
 \right. \non && \left.
\hspace{0.3cm}\times
 E_{fa}(t_{b})\; h_{fa}(x_{2},1-x_{3},b_{2},b_{3}) \right\}\;,
 \label{eq:abt-aa}
\eeq
where the superscripts $V$ and $A$ in the formulas express
the types of mesons involved in the considered decays, the
subscripts ${fa}$ and ${na}$ (to be shown below) are the
abbreviations of factorizable annihilation and nonfactorizable annihilation
 respectively, and $C_F=4/3$ is a color factor.
Moreover, the terms proportional to $r_{2(3)}^2$ can not change the results significantly
and they have been neglected safely
because the values of $r_{2(3)}^2$ are numerically small: $r_{2(3)}^2 < 5\%$.  For
the function $h_{fa}$, the scales $t_i$, and $E_{fa}(t)$, we use
the expressions as given in Appendix~B of Ref.~\cite{xiao09:bcnd}.

For the nonfactorizable diagrams (c) and (d) in Fig.~\ref{fig:fig1}, all three meson wave
functions are involved. The integration of $b_3$ can be performed
using $\delta$ function $\delta(b_3-b_2)$, leaving only integration
of $b_1$ and $b_2$. The corresponding decay amplitudes are
\beq
M_{na}^{L}(AV) &=& -\frac{16 \sqrt{6}}{3}\pi C_F m_{B_c}^2 \int_{0}^{1}d x_{2}\,d x_{3}\,\int_{0}^{\infty} b_1 db_1 b_2 db_2\,
 \non && \times
 \left\{E_{na}(t_c)\left[ (r_c - x_3 +1) \phi_{A}(x_2)\phi_{V}(x_{3}) - r_2^A r_3^V \left(\phi_{A}^s(x_2)((3 r_c + x_2 -x_3 +1)
  \right.\right.\right. \non && \left. \left. \left. \times
 \phi_{V}^s(x_3)-(r_c -x_2 -x_3 +1) \phi_{V}^t(x_3))+\phi_{A}^t(x_2)((r_c-x_2 -x_3 +1) \phi_{V}^s(x_3)
 \right.\right.\right. \non && \left.\left. \left.
+(r_c -x_2 +x_3 -1)\phi_{V}^t(x_3))\right)\right] h_{na}^{c}(x_2,x_3,b_1,b_2)-h_{na}^{d}(x_2,x_3,b_1,b_2)E_{na}(t_d)
\right. \non && \left.  \times
\left[ (r_b + r_c +x_2 -1) \phi_{A}(x_2) \phi _{V}(x_3) - r_2^A r_3^V \left(\phi_{A}^s(x_2)((4 r_b+r_c +x_2 -x_3 -1)
\right.\right.\right.\non && \left. \left.\left. \times
\phi_{V}^s(x_3)-(r_c + x_2 +x_3 -1)\phi_{V}^t(x_3))+\phi_{A}^t(x_2)((r_c + x_2 +x_3 -1)\phi_{V}^s(x_3)
\right.\right.\right. \non && \left. \left. \left.
 -(r_c +x_2 -x_3 -1) \phi_{V}^t(x_3))\right)\right] \right\}\;,\label{eq:cdl-av}\\
M_{na}^{N}(AV) &=& -\frac{32 \sqrt{6}}{3}\pi C_F m_{B_c}^2 \int_{0}^{1}d x_{2}\,d x_{3}\,\int_{0}^{\infty} b_1d b_1 b_2db_2\, r_2^A r_3^V
 \non && \times
 \left\{r_c\left[ \phi_{A}^a(x_2)\phi_{V}^a(x_{3})+\phi_{A}^v(x_2)\phi_{V}^v (x_{3}) \right]E_{na}(t_c) h_{na}^{c}(x_2,x_3,b_1,b_2)
  \right. \non && \left.
  -r_b\left[ \phi_{A}^a(x_2)\phi_{A}^a(x_{3})+\phi_{A}^v(x_2)\phi_{V}^v (x_{3})\right] E_{na}(t_d)h_{na}^{d}(x_2,x_3,b_1,b_2)\right\}\;,
  \label{eq:cdn-av}\eeq
  \beq
 M_{na}^{T}(AV) &=& -\frac{64 \sqrt{6}}{3}\pi C_F m_{B_c}^2 \int_{0}^{1}d x_{2}\,d x_{3}\,\int_{0}^{\infty} b_1d b_1 b_2db_2\, r_2^A r_3^V
 \non && \times
 \left\{r_c\left[ \phi_{A}^a(x_2)\phi_{V}^v(x_{3})+\phi_{A}^v(x_2)\phi_{V}^a (x_{3}) \right]E_{na}(t_c) h_{na}^{c}(x_2,x_3,b_1,b_2)
 \right. \non && \left.
  -r_b\left[ \phi_{A}^a(x_2)\phi_{V}^v(x_{3})+\phi_{A}^v(x_2)\phi_{V}^a (x_{3}) \right] E_{na}(t_d)h_{na}^{d}(x_2,x_3,b_1,b_2)\right\}\;.
  \label{eq:cdt-av}
  \eeq
\beq
M_{na}^{L}(VA) &=& -\frac{16 \sqrt{6}}{3}\pi C_F m_{B_c}^2 \int_{0}^{1}d x_{2}\,d x_{3}\,\int_{0}^{\infty} b_1d b_1 b_2 db_2\,
 \non && \times
 \left\{E_{na}(t_c)\left[ (r_c - x_3 +1) \phi_{V}(x_2)\phi_{A}(x_{3}) + r_2^V r_3^A \left(\phi_{V}^s(x_2)((3 r_c + x_2 -x_3 +1)
  \right.\right.\right. \non && \left. \left. \left. \times
 \phi_{A}^s(x_3)-(r_c -x_2 -x_3 +1) \phi_{A}^t(x_3))+\phi_{V}^t(x_2)((r_c-x_2 -x_3 +1) \phi_{A}^s(x_3)
 \right.\right.\right. \non && \left.\left. \left.
+(r_c -x_2 +x_3 -1)\phi_{A}^t(x_3))\right)\right] h_{na}^{c}(x_2,x_3,b_1,b_2)-h_{na}^{d}(x_2,x_3,b_1,b_2)E_{na}(t_d)
\right. \non && \left.  \times
\left[ (r_b + r_c +x_2 -1) \phi_{V}(x_2) \phi _{A}(x_3) + r_2^V r_3^A \left(\phi_{V}^s(x_2)((4 r_b+r_c +x_2 -x_3 -1)
\right.\right.\right.\non && \left. \left.\left. \times
 \phi_{A}^s(x_3)-(r_c + x_2 +x_3 -1)\phi_{A}^T(x_3))+\phi_{V}^t(x_2)((r_c + x_2 +x_3 -1)\phi_{A}^s(x_3)
 \right.\right.\right. \non && \left. \left. \left.
 -(r_c +x_2 -x_3 -1) \phi_{A}^t(x_3))\right)\right] \right\}\;,\label{eq:cdl-va}\\
M_{na}^{N}(VA) &=& -\frac{32 \sqrt{6}}{3}\pi C_F m_{B_c}^2 \int_{0}^{1}d x_{2}\,d x_{3}\,\int_{0}^{\infty} b_1d b_1 b_2d b_2\, r_2^V r_3^A
 \non && \times
 \left\{r_c\left[ \phi_{V}^a(x_2)\phi_{A}^a(x_{3})+\phi_{V}^v(x_2)\phi_{A}^v (x_{3}) \right]E_{na}(t_c) h_{na}^{c}(x_2,x_3,b_1,b_2)
  \right. \non && \left.
  -r_b\left[ \phi_{V}^a(x_2)\phi_{A}^a(x_{3})+\phi_{V}^v(x_2)\phi_{A}^v (x_{3}) \right] E_{na}(t_d)h_{na}^{d}(x_2,x_3,b_1,b_2)\right\}\;,
  \label{eq:cdn-va}\\
M_{na}^{T}(VA) &=& - \frac{64 \sqrt{6}}{3}\pi C_F m_{B_c}^2 \int_{0}^{1}d x_{2}\,d x_{3}\,\int_{0}^{\infty} b_1d b_1 b_2d b_2\, r_2^V r_3^A
 \non && \times
 \left\{r_c\left[ \phi_{V}^a(x_2)\phi_{A}^v(x_{3})+\phi_{V}^v(x_2)\phi_{A}^a (x_{3}) \right]E_{na}(t_c) h_{na}^{c}(x_2,x_3,b_1,b_2)
 \right. \non && \left.
 -r_b\left[ \phi_{V}^a(x_2)\phi_{A}^v(x_{3})+\phi_{V}^v(x_2)\phi_{A}^a (x_{3}) \right] E_{na}(t_d)h_{na}^{d}(x_2,x_3,b_1,b_2)\right\}\;
.\label{eq:cdt-va}
\eeq
\beq
M_{na}^{L}(AA) &=& \frac{16 \sqrt{6}}{3}\pi C_F m_{B_c}^2 \int_{0}^{1}d x_{2}\,d x_{3}\,\int_{0}^{\infty} b_1 db_1 b_2 db_2\,
 \non && \times
\left\{E_{na}(t_c)\left[ (r_c - x_3 +1)\phi_{2}(x_2)\phi_{3}(x_{3}) + r_2^A r_3^A \left(\phi_{2}^s(x_2)
((3 r_c + x_2 -x_3 +1)  \right.\right.\right. \non && \left. \left. \left. \times \phi_{3}^s(x_3)-(r_c -x_2 -x_3 +1)
\phi_{3}^t(x_3))+\phi_{2}^t(x_2)((r_c-x_2 -x_3 +1) \phi_{3}^s(x_3)\right.\right.\right. \non && \left.\left. \left.
+(r_c -x_2 +x_3 -1)\phi_{3}^t(x_3))\right)\right] h_{na}^{c}(x_2,x_3,b_1,b_2)-h_{na}^{d}(x_2,x_3,b_1,b_2)E_{na}(t_d)
\right. \non && \left.  \times
\left[ (r_b + r_c +x_2 -1) \phi_{2}(x_2) \phi _{3}(x_3) + r_2^A r_3^A \left(\phi_{2}^s(x_2)((4 r_b+r_c +x_2 -x_3 -1)
\right.\right.\right.\non && \left. \left.\left. \times
\phi_{3}^s(x_3)-(r_c + x_2 +x_3 -1)\phi_{3}^t(x_3))+\phi_{2}^t(x_2)((r_c + x_2 +x_3 -1)\phi_{3}^s(x_3)
\right.\right.\right. \non && \left. \left.\left.
 -(r_c +x_2 -x_3 -1) \phi_{3}^t(x_3))\right)\right]\right\}\;,\label{eq:cdl-aa}\\
M_{na}^{N}(AA) &=& \frac{32 \sqrt{6}}{3}\pi C_F m_{B_c}^2 \int_{0}^{1}d x_{2}\,d x_{3}\,\int_{0}^{\infty}
 b_1 db_1 b_2 db_2\, r_2^A r_3^A \non && \times
 \left\{r_c\left[ \phi_{2}^a(x_2)\phi_{3}^a(x_{3})+\phi_{2}^v(x_2)\phi_{3}^v (x_{3})
 \right]E_{na}(t_c) h_{na}^{c}(x_2,x_3,b_1,b_2) \right. \non &&
\left.  -r_b\left[ \phi_{2}^a(x_2)\phi_{3}^a(x_{3})+\phi_{2}^v(x_2)\phi_{3}^v (x_{3})
\right]E_{na}(t_d)h_{na}^{d}(x_2,x_3,b_1,b_2)\right\}\;,\label{eq:cdn-aa}\\
M_{na}^{T}(AA) &=& \frac{64 \sqrt{6}}{3}\pi C_F m_{B_c}^2 \int_{0}^{1}d x_{2}\,d x_{3}\,\int_{0}^{\infty} b_1d b_1 b_2db_2\, r_2^A r_3^A
 \non && \times
 \left\{r_c\left[ \phi_{2}^a(x_2)\phi_{3}^v(x_{3})+\phi_{2}^v(x_2)\phi_{3}^a (x_{3})
 \right]E_{na}(t_c) h_{na}^{c}(x_2,x_3,b_1,b_2)
\right. \non && \left.
-r_b\left[ \phi_{2}^a(x_2)\phi_{3}^v(x_{3})+\phi_{2}^v(x_2)\phi_{3}^a (x_{3})
\right] E_{na}(t_d)h_{na}^{d}(x_2,x_3,b_1,b_2)\right\}\;
.\label{eq:cdt-aa}
\eeq
where $r_{b}= m_{b}/m_{B_c}$, $r_{c}= m_{c}/m_{B_c}$,  and $r_b +r_c \approx 1$ for $B_c$ meson.

There are three kinds of polarizations in these $B_c \to VA, AA$ decays, namely, longitudinal ($L$), normal ($N$) and
transverse ($T$). The decay amplitudes are classified accordingly, with $H=L,N,T$.
From the effective Hamiltonian~(\ref{heff}), based on
Eqs.~(\ref{eq:abl-av}-\ref{eq:cdt-aa}), we can combine all
contributions to these considered decays and obtain the total decay amplitude
generally as,
\beq
{\cal M}^H(B_c \to  M_2 M_3) &=& V_{cb}^* V_{uD}
\left\{f_{B_c} F_{fa;H}^{M_2 M_3}a_1 + M_{na;H}^{M_2 M_3} C_1 \right\}
\; , \label{eq:amt}
\eeq
where $a_1=C_1/3+C_2$. Then we can write
down the total decay amplitudes for sixty two charmless two-body nonleptonic $B_c$ meson decays into
final states involving one vector and one axial-vector meson ($VA$)
or two axial-vector mesons($AA$) one by one.

\vspace{5.0mm}

1.  $B_c \to VA/AV$ decay modes

\begin{itemize}

\item[]{(i)} For $\Delta S =0$ processes,
\beq
\sqrt{2}{\cal M}^{H}(B_c \to \rho^+  a_1^0)&=&  V_{cb}^* V_{ud} \left\{\left [f_{B_c} F_{fa;H}^{\rho a_{1u}^0}a_1
+ M_{na;H}^{\rho a_{1u}^0} C_1 \right ]
  \right. \non && \left.
-\left [f_{B_c}F_{fa;H}^{a_{1d}^0\rho}a_1 + M_{na;H}^{a_{1d}^0\rho }
C_1 \right ] \right\}\;,\label{eq:rhopa1} \\
\sqrt{2}{\cal M}^{H}(B_c \to a_1^+ \rho^0) &=&  - \sqrt{2}{\cal M}^{H}(B_c \to \rho^+  a_1^0)
\non
&=& V_{cb}^* V_{ud} \left\{\left [f_{B_c} F_{fa;H}^{a_1^+ \rho_{u}^0 }
a_1 + M_{na;H}^{a_1^+ \rho_{u}^0} C_1\right ]
 \right. \non && \left.
- \left [f_{B_c} F_{fa;H}^{\rho_{d}^0 a_1^+ }a_1 + M_{na;H}^{\rho_{d}^0 a_1^+ }
C_1\right ]\right\} \;,\label{eq:rhoa1p}\\
\sqrt{2}{\cal M}^{H}(B_c \to a_1^+ \omega) &=& V_{cb}^* V_{ud}
\left\{\left [f_{B_c} F_{fa;H}^{a_1^+ \omega_{u} }a_1 + M_{na;H}^{a_1^+ \omega_{u}} C_1\right ]
 \right. \non && \left.
+ \left [f_{B_c} F_{fa;H}^{\omega_{d} a_1^+ }a_1 + M_{na;H}^{\omega_{d} a_1^+ } C_1\right ]
\right\}\;,\label{eq:a1pome}
\eeq
\beq
\sqrt{2}{\cal M}^{H}(B_c \to \rho^+  b_1^0)&=&  V_{cb}^* V_{ud} \left\{\left [f_{B_c} F_{fa;H}^{\rho
b_{1u}^0}a_1 + M_{na;H}^{\rho b_{1u}^0} C_1 \right ]
  \right. \non && \left.
-\left [f_{B_c}F_{fa;H}^{b_{1d}^0 \rho}a_1 + M_{na;H}^{b_{1d}^0\rho } C_1 \right ]
\right\}\;,\label{eq:rhopb1} \\
\sqrt{2}{\cal M}^{H}(B_c \to b_1^+ \rho^0) &=&  - \sqrt{2}{\cal M}^{H}(B_c \to \rho^+  b_1^0)
\non
&=& V_{cb}^* V_{ud} \left\{\left [f_{B_c} F_{fa;H}^{b_1^+
\rho_{u}^0 }a_1 + M_{na;H}^{b_1^+ \rho_{u}^0} C_1\right ]
 \right. \non && \left.
- \left [f_{B_c} F_{fa;H}^{\rho_{d}^0 b_1^+ }a_1 + M_{na;H}^{\rho_{d}^0 b_1^+ }
C_1\right ]\right\}\;,\label{eq:b1prho}\\
\sqrt{2}{\cal M}^{H}(B_c \to b_1^+ \omega) &=& V_{cb}^* V_{ud}  \left\{\left [f_{B_c} F_{fa;H}^{b_1^+
\omega_{u} }a_1 + M_{na;H}^{b_1^+ \omega_{u}} C_1\right ]
 \right. \non && \left.
+ \left [f_{B_c} F_{fa;H}^{\omega_{d} b_1^+ }a_1 + M_{na;H}^{\omega_{d} b_1^+}
C_1\right ]\right\}\;,\label{eq:b1pome}
\eeq
\beq
{\cal M}^{H}(B_c \to \rho^+  f')&=&  V_{cb}^* V_{ud}
\left\{\frac{\cos\theta_{3}}{\sqrt{3}}
\left [f_{B_c} (F_{fa;H}^{\rho
f_{1}^{u}}+F_{fa;H}^{f_{1}^{d}\rho})a_1
\right. \right.\non && \left.\left.
+ (M_{na;H}^{\rho f_{1}^{u}}+M_{na;H}^{f_{1}^{d}\rho}) C_1 \right ]
 +\frac{\sin\theta_{3}}{\sqrt{6}}\left [f_{B_c}
\right. \right.\non && \left.\left.
\cdot (F_{fa;H}^{\rho f_{8}^{u}}+F_{fa;H}^{f_{8}^{d}\rho}) a_1 + (M_{na;H}^{\rho
f_{8}^{u}}+M_{na;H}^{f_{8}^{d}\rho }) C_1 \right ] \right\}\;,\label{eq:rhopf'}
\eeq
\beq
{\cal M}^{H}(B_c \to \rho^+  f^{''})&=&  V_{cb}^* V_{ud}
\left\{\frac{-\sin\theta_{3}}{ \sqrt{3}}\left [f_{B_c} (F_{fa;H}^{\rho
f_{1}^{u}}+F_{fa;H}^{f_{1}^{d}\rho})a_1
 \right. \right. \non && \left.\left.
+ (M_{na;H}^{\rho f_{1}^{u}}+M_{na;H}^{f_{1}^{d}\rho}) C_1 \right ]
 +\frac{\cos\theta_{3}}{\sqrt{6}}\left [f_{B_c}
\right. \right.\non && \left. \left.
\cdot (F_{fa;H}^{\rho f_{8}^{u}}+F_{fa;H}^{f_{8}^{d}\rho}) a_1
+ (M_{na;H}^{\rho f_{8}^{u}}+M_{na;H}^{f_{8}^{d}\rho}) C_1 \right ] \right\}\;,\label{eq:rhopf''}
\eeq
\beq
{\cal M}^{H}(B_c \to \rho^+  h')&=&  V_{cb}^* V_{ud}
\left\{\frac{\cos\theta_{1}}{\sqrt{3}}\left [f_{B_c} (F_{fa;H}^{\rho
h_{1}^{u}}+F_{fa;H}^{h_{1}^{d}\rho})a_1
\right.\right. \non && \left.\left.
+ (M_{na;H}^{\rho h_{1}^{u}}+M_{na;H}^{h_{1}^{d}\rho}) C_1 \right ]
 +\frac{\sin\theta_{1}}{\sqrt{6}}\left [f_{B_c}
\right.\right. \non && \left. \left.
\cdot (F_{fa;H}^{\rho h_{8}^{u}}+F_{fa;H}^{h_{8}^{d}\rho}) a_1 + (M_{na;H}^{\rho h_{8}^{u}}
+M_{na;H}^{h_{8}^{d}\rho }) C_1 \right ] \right\}\;,\label{eq:rhoph'}
\eeq
\beq
{\cal M}^{H}(B_c \to \rho^+  h^{''})&=&  V_{cb}^* V_{ud}
\left\{\frac{-\sin\theta_{1}}{\sqrt{3}}\left [f_{B_c} (F_{fa;H}^{\rho
h_{1}^{u}}+F_{fa;H}^{h_{1}^{d}\rho})a_1
\right. \right. \non && \left. \left.
+ (M_{na;H}^{\rho h_{1}^{u}}+M_{na;H}^{h_{1}^{d}\rho}) C_1 \right ]
 +\frac{\cos\theta_{1}}{\sqrt{6}}\left [f_{B_c}
\right.\right. \non && \left.\left.
\cdot (F_{fa;H}^{\rho h_{8}^{u}}+F_{fa;H}^{h_{8}^{d}\rho}) a_1 + (M_{na;H}^{\rho h_{8}^{u}}
+M_{na;H}^{h_{8}^{d}\rho }) C_1 \right ] \right\}\;,\label{eq:rhoph''}
\eeq
\beq
{\cal M}^{H}(B_c \to \ov{K^{*0}} {K'}^+) &=& V_{cb}^* V_{ud}\left\{\sin \theta_{K}
\left [f_{B_c} F_{fa;H}^{\ov{K^{*0}} K_{1A}}a_1 +
M_{na;H}^{\ov{K^{*0}} K_{1A}} C_1 \right ]
\right. \non && \left.
+\cos \theta_{K} \left [f_{B_c} F_{fa;H}^{\ov{K^{*0}} K_{1B}}a_1 +
M_{na;H}^{\ov{K^{*0}} K_{1B}} C_1 \right ] \right\}\;,\label{eq:kstk127}\\
{\cal M}^{H}(B_c \to \ov{K^{*0}} {K^{''}}^+) &=& V_{cb}^* V_{ud}\left\{\cos \theta_{K}
 \left [f_{B_c} F_{fa;H}^{\ov{K^{*0}} K_{1A}}a_1 +
M_{na;H}^{\ov{K^{*0}} K_{1A}} C_1 \right ]
\right. \non && \left.
-\sin \theta_{K}\left [f_{B_c} F_{fa;H}^{\ov{K^{*0}} K_{1B}}a_1 +
M_{na;H}^{\ov{K^{*0}} K_{1B}} C_1 \right ] \right\}\;,\label{eq:kstk140}
\eeq
\beq
{\cal M}^{H}(B_c \to \ov{K'}^0 {K^{*}}^+) &=& V_{cb}^* V_{ud}\left\{\sin \theta_{K}
\left [f_{B_c} F_{fa;H}^{\ov{K_{1A}^0} K^*}a_1 +
M_{na;H}^{\ov{K_{1A}^0} K^*} C_1\right ]
\right. \non && \left.
+\cos \theta_{K}\left [f_{B_c} F_{fa;H}^{\ov{K_{1B}^0} K^*}a_1 +
M_{na;H}^{\ov{K_{1B}^0} K^*} C_1 \right ] \right\}\;,\label{eq:k127kst}\\
{\cal M}^{H}(B_c \to \ov{K^{''}}^0 {K^{*}}^+) &=& V_{cb}^* V_{ud}\left\{\cos \theta_{K}
\left [f_{B_c} F_{fa;H}^{\ov{K_{1A}^0} K^*}a_1 +
M_{na;H}^{\ov{K_{1A}^0} K^*} C_1\right ]
\right. \non && \left.
-\sin \theta_{K}\left [f_{B_c} F_{fa;H}^{\ov{K_{1B}^0} K^*}a_1 +
M_{na;H}^{\ov{K_{1B}^0} K^*} C_1 \right ] \right\}\;. \label{eq:k140kst}
\eeq


\item[]{(ii)} For $\Delta S =1$ processes,
\beq
{\cal M}^{H}(B_c \to  {K^{*}}^0 a_1^+) &=&  \sqrt{2}{\cal M}^{H}(B_c \to  {K^{*}}^+ a_1^0)
\non
&=& V_{cb}^* V_{us} \left\{f_{B_c}
F_{fa;H}^{K^{*0} a_1^+} a_1 + M_{na;H}^{K^{*0} a_1^+} C_1 \right\},\; \label{eq:ksta1p}\\
{\cal M}^{H}(B_c \to  {K^{*}}^0 b_1^+) &=&  \sqrt{2}{\cal M}^{H}(B_c \to  {K^{*}}^+ b_1^0)
\non
&=& V_{cb}^* V_{us} \left\{f_{B_c}
F_{fa;H}^{K^{*0} b_1^+} a_1 + M_{na;H}^{K^{*0} b_1^+} C_1 \right\},\;\label{eq:kstb1p}
\eeq
\beq
{\cal M}^{H}(B_c \to  {K'}^0 \rho^+) &=& \sqrt{2} {\cal M}^{H}(B_c \to  {K'}^+ \rho^0)\non
&=& V_{cb}^* V_{us} \left\{\sin \theta_{K}\left [f_{B_c}
F_{fa;H}^{K_{1A}^{0}\rho}a_1 + M_{na;H}^{K_{1A}^{0}\rho} C_1\right ]
\right. \non && \left.
+ \cos \theta_{K} \left [f_{B_c}
F_{fa;H}^{K_{1B}^{0}\rho}a_1 + M_{na;H}^{K_{1B}^{0}\rho} C_1\right ] \right\},
\label{eq:k127rho}
\eeq
\beq
{\cal M}^{H}(B_c \to  {K^{''}}^0 \rho^+) &=& \sqrt{2} {\cal M}^{H}(B_c \to  {K^{''}}^+ \rho^0)
\non
&=& V_{cb}^* V_{us} \left\{\cos \theta_{K}\left [f_{B_c}
F_{fa;H}^{K_{1A}^{0}\rho}a_1 + M_{na;H}^{K_{1A}^{0}\rho} C_1\right ]
\right. \non && \left.
- \sin \theta_{K} \left [f_{B_c}
F_{fa;H}^{K_{1B}^{0}\rho}a_1 + M_{na;H}^{K_{1B}^{0}\rho} C_1\right ] \right\},
\label{eq:k140rho}
\eeq
\beq
 \sqrt{2} {\cal M}^{H}(B_c \to  K^{'+} \omega)
&=& V_{cb}^* V_{us} \left\{\sin \theta_{K}\left [f_{B_c}
F_{fa;H}^{K_{1A}^{0}\omega}a_1 + M_{na;H}^{K_{1A}^{0}\omega} C_1\right ]
\right. \non && \left.
+ \cos \theta_{K} \left [f_{B_c}
F_{fa;H}^{K_{1B}^{0}\omega}a_1 + M_{na;H}^{K_{1B}^{0}\omega} C_1\right ] \right\},
\label{eq:k127ome}\\
 \sqrt{2} {\cal M}^{H}(B_c \to  {K^{''}}^+ \omega)
&=& V_{cb}^* V_{us} \left\{\cos \theta_{K}\left [f_{B_c}
F_{fa;H}^{K_{1A}^{0}\omega}a_1 + M_{na;H}^{K_{1A}^{0}\omega} C_1\right ]
\right. \non && \left.
- \sin \theta_{K} \left [f_{B_c}
F_{fa;H}^{K_{1B}^{0}\omega}a_1 + M_{na;H}^{K_{1B}^{0}\omega} C_1\right ] \right\},
\label{eq:k140ome}
\eeq
\beq
{\cal M}^{H}(B_c \to {K^{*}}^+  f')&=&
V_{cb}^* V_{us} \left\{ \frac{\cos\theta_{3}}{\sqrt{3}}
\left [f_{B_c} (F_{fa;H}^{K^*
f_{1}^{u}}+F_{fa;H}^{f_{1}^{s} K^*})a_1
\right. \right.\non && \left.\left.
+ (M_{na;H}^{K^* f_{1}^{u}}+M_{na;H}^{f_{1}^{s} K^*}) C_1 \right ]
 +\frac{\sin\theta_{3}}{\sqrt{6}}\left [f_{B_c}
\right.\right. \non && \left.\left.
\cdot (F_{fa;H}^{K^* f_{8}^{u}}-2 F_{fa;H}^{f_{8}^{s} K^*}) a_1 + (M_{na;H}^{K^* f_{8}^{u}}
-2 M_{na;H}^{f_{8}^{s} K^* }) C_1 \right ] \right\}\;,\label{eq:kstf1}
\eeq
\beq
{\cal M}^{H}(B_c \to {K^{*}}^+  f^{''})&=&  V_{cb}^* V_{us} \left\{\frac{-\sin\theta_{3}}{
\sqrt{3}}\left [f_{B_c} (F_{fa;H}^{K^*
f_{1}^{u}}+F_{fa;H}^{f_{1}^{s} K^*})a_1
\right.\right. \non && \left.\left.
+ (M_{na;H}^{K^* f_{1}^{u}}+M_{na;H}^{f_{1}^{s} K^*}) C_1 \right ]
 +\frac{\cos\theta_{3}}{\sqrt{6}}\left [f_{B_c}
\right.\right. \non && \left.\left.
\cdot (F_{fa;H}^{K^* f_{8}^{u}}- 2 F_{fa;H}^{f_{8}^{s} K^*}) a_1 + (M_{na;H}^{K^* f_{8}^{u}}
-2 M_{na;H}^{f_{8}^{s} K^* }) C_1 \right ] \right\}\;,\label{eq:kstf2}
\eeq
\beq
{\cal M}^{H}(B_c \to {K^{*}}^+  h')&=&  V_{cb}^* V_{us} \left\{\frac{\cos\theta_{1}}{\sqrt{3}}
\left [f_{B_c} (F_{fa;H}^{K^* h_{1}^{u}}+F_{fa;H}^{h_{1}^{s} K^*})a_1
\right.\right. \non && \left.\left.
+ (M_{na;H}^{K^* h_{1}^{u}}+M_{na;H}^{h_{1}^{s} K^*}) C_1 \right ]
 +\frac{\sin\theta_{1}}{\sqrt{6}}\left [f_{B_c}
\right.\right. \non && \left.\left.
\cdot (F_{fa;H}^{K^* h_{8}^{u}}-2 F_{fa;H}^{h_{8}^{s} K^*}) a_1 + (M_{na;H}^{K^* h_{8}^{u}}
-2 M_{na;H}^{h_{8}^{s} K^* }) C_1 \right ] \right\}\;,\label{eq:kstf3}
\eeq
\beq
{\cal M}^{H}(B_c \to {K^{*}}^+  h^{''})&=&  V_{cb}^* V_{us}
\left\{\frac{-\sin\theta_{1}}{\sqrt{3}}
\left [f_{B_c} (F_{fa;H}^{K^* h_{1}^{u}}+F_{fa;H}^{h_{1}^{s} K^*})a_1
\right. \right. \non && \left. \left.
+ (M_{na;H}^{K^* h_{1}^{u}}+M_{na;H}^{h_{1}^{s} K^*}) C_1 \right ]
 +\frac{\cos\theta_{1}}{\sqrt{6}}\left [f_{B_c}
\right.\right. \non && \left.\left.
\cdot (F_{fa;H}^{K^* h_{8}^{u}}-2 F_{fa;H}^{h_{8}^{s} K^*}) a_1 + (M_{na;H}^{K^* h_{8}^{u}}
-2 M_{na;H}^{h_{8}^{s} K^* }) C_1 \right ] \right\}\;,\label{eq:kstf4}
\eeq
\beq
 {\cal M}^{H}(B_c \to \phi K^{'+} )
&=& V_{cb}^* V_{us} \left\{\sin \theta_{K}\left [f_{B_c}
F_{fa;H}^{\phi K_{1A}^{0}}a_1 + M_{na;H}^{\phi K_{1A}^{0}} C_1\right ]
\right. \non && \left.
+ \cos \theta_{K} \left [f_{B_c}
F_{fa;H}^{\phi K_{1B}^{0}}a_1 + M_{na;H}^{\phi K_{1B}^{0}} C_1\right ] \right\},
\label{eq:phik127}\\
{\cal M}^{H}(B_c \to  \phi {K^{''}}^+)
&=& V_{cb}^* V_{us} \left\{\cos \theta_{K}\left [f_{B_c}
F_{fa;H}^{\phi K_{1A}^{0}}a_1 + M_{na;H}^{\phi K_{1A}^{0}} C_1\right ]
\right. \non && \left.
- \sin \theta_{K} \left [f_{B_c}
F_{fa;H}^{\phi K_{1B}^{0}}a_1 + M_{na;H}^{\phi K_{1B}^{0}} C_1\right ] \right\}.
\label{eq:phik140}
\eeq

\end{itemize}

\vspace{5.0mm}

2.  $B_c \to AA$ decay modes

\begin{itemize}

\item[]{(i)} For $\Delta S= 0$ processes,
\beq
\sqrt{2} {\cal M}^{H}(B_c \to a_1^+  a_1^0)&=&  V_{cb}^* V_{ud} \left\{f_{B_c}
(F_{fa;H}^{a_1^+ a_{1u}^0}-F_{fa;H}^{a_{1d}^0 a_1^+} )a_1
\right. \non && \left.
+ (M_{na;H}^{a_1^+ a_{1u}^0}-M_{na;H}^{a_{1d}^0 a_1^+ }) C_1 \right\}\;,\label{eq:a1pa10} \\
\sqrt{2} {\cal M}^{H}(B_c \to b_1^+  b_1^0)&=&  V_{cb}^* V_{ud} \left\{f_{B_c}
(F_{fa;H}^{b_1^+ b_{1u}^0}-F_{fa;H}^{b_{1d}^0 b_1^+})a_1
\right. \non && \left.
 + (M_{na;H}^{b_1^+ b_{1u}^0}- M_{na;H}^{b_{1d}^0 b_1^+ }) C_1 \right\} \;,\label{eq:b1pb10}
 \eeq
\beq
\sqrt{2} {\cal M}^{H}(B_c \to a_1^+  b_1^0)&=&  V_{cb}^* V_{ud} \left\{f_{B_c}
(F_{fa;H}^{a_1^+ b_{1u}^0}- F_{fa;H}^{b_{1d}^0 a_1^+})a_1
\right. \non && \left.
+ (M_{na;H}^{a_1^+ b_{1u}^0}- M_{na;H}^{b_{1d}^0 a_1^+ }) C_1 \right\} \;,\label{eq:a1pb10} \\
\sqrt{2} {\cal M}^{H}(B_c \to b_1^+  a_1^0)&=&  V_{cb}^* V_{ud} \left\{ f_{B_c}
(F_{fa;H}^{b_1^+ a_{1u}^0}- F_{fa;H}^{a_{1d}^0 b_1^+})a_1
\right. \non && \left.
 + (M_{na;H}^{b_1^+ a_{1u}^0}- M_{na;H}^{a_{1d}^0 b_1^+}) C_1  \right\} \;,\label{eq:b1pa10}
 \eeq
\beq
{\cal M}^{H}(B_c \to a_1^+  f')&=&  V_{cb}^* V_{ud} \left\{ \frac{\cos\theta_{3}}{\sqrt{3}}[f_{B_c}
(F_{fa;H}^{a_1^+ f_{1}^u}+F_{fa;H}^{f_{1}^d a_1^+})a_1
\right. \non && \left.
 + (M_{na;H}^{a_1^+ f_{1}^u}+M_{na;H}^{f_{1}^d a_1^+}) C_1 ]+ \frac{\sin\theta_{3}}{\sqrt{6}}[f_{B_c}
 \right. \non && \left.
\cdot (F_{fa;H}^{a_1^+ f_{8}^u}+F_{fa;H}^{f_{8}^d a_1^+}) a_1
 + (M_{na;H}^{a_1^+ f_{8}^u}+M_{na;H}^{f_{8}^d a_1^+ }) C_1 ] \right\}\;,\label{eq:a1pf1'}
 \eeq
\beq
{\cal M}^{H}(B_c \to a_1^+  f^{''})&=&  V_{cb}^* V_{ud} \left\{\frac{-\sin\theta_{3}}{\sqrt{3}}[f_{B_c} (F_{fa;H}^{a_1^+
f_{1}^u}+F_{fa;H}^{f_{1}^d a_1^+})a_1
\right. \non && \left.
 + (M_{na;H}^{a_1^+ f_{1}^u}+M_{na;H}^{f_{1}^d a_1^+}) C_1 ] +\frac{\cos\theta_{3}}{\sqrt{6}}[f_{B_c}
\right. \non && \left.
\cdot (F_{fa;H}^{a_1^+ f_{8}^u}+F_{fa;H}^{f_{8}^d a_1^+}) a_1
+ (M_{na;H}^{a_1^+ f_{8}^u}+M_{na;H}^{f_{8}^d a_1^+ }) C_1 ] \right\}\;,\label{eq:a1pf1''}
\eeq
\beq
{\cal M}^{H}(B_c \to b_1^+  f')&=&  V_{cb}^* V_{ud} \left\{ \frac{\cos\theta_{3}}{\sqrt{3}}[f_{B_c} (F_{fa;H}^{b_1^+
f_{1}^u}+F_{fa;H}^{f_{1}^d b_1^+})a_1
\right. \non && \left.
+ (M_{na;H}^{b_1^+ f_{1}^u}+M_{na;H}^{f_{1}^d b_1^+}) C_1 ] +\frac{\sin\theta_{3}}{\sqrt{6}}[f_{B_c}
\right. \non && \left.
\cdot (F_{fa;H}^{b_1^+ f_{8}^u}+F_{fa;H}^{f_{8}^d b_1^+}) a_1
 + (M_{na;H}^{b_1^+ f_{8}^u}+M_{na;H}^{f_{8}^d b_1^+ }) C_1 ] \right\}\;,\label{eq:b1pf1'}
\eeq
\beq
{\cal M}^{H}(B_c \to b_1^+  f^{''})&=&  V_{cb}^* V_{ud} \left\{\frac{- \sin\theta_{3}}{\sqrt{3}}[f_{B_c} (F_{fa;H}^{b_1^+
f_{1}^u}+F_{fa;H}^{f_{1}^d b_1^+})a_1
\right. \non && \left.
 + (M_{na;H}^{b_1^+ f_{1}^u}+M_{na;H}^{f_{1}^d b_1^+}) C_1 ] +\frac{\cos\theta_{3}}{\sqrt{6}}[f_{B_c}
 \right. \non && \left.
\cdot (F_{fa;H}^{b_1^+ f_{8}^u}+F_{fa;H}^{f_{8}^d b_1^+}) a_1
+ (M_{na;H}^{b_1^+ f_{8}^u}+M_{na;H}^{f_{8}^d b_1^+ }) C_1 ] \right\}\;,\label{eq:b1pf1''}
\eeq
\beq
{\cal M}^{H}(B_c \to a_1^+  h')&=&  V_{cb}^* V_{ud} \left\{  \frac{\cos\theta_{1}}{\sqrt{3}}[f_{B_c} (F_{fa;H}^{a_1^+
h_{1}^u}+F_{fa;H}^{h_{1}^d a_1^+})a_1
\right.\non && \left.
+ (M_{na;H}^{a_1^+ h_{1}^u}+M_{na;H}^{h_{1}^d a_1^+}) C_1 ] + \frac{\sin\theta_{1}}{\sqrt{6}}[f_{B_c}
\right.\non && \left.
\cdot (F_{fa;H}^{a_1^+ h_{8}^u}+F_{fa;H}^{h_{8}^d a_1^+}) a_1
 + (M_{na;H}^{a_1^+ h_{8}^u}+M_{na;H}^{h_{8}^d a_1^+ }) C_1 ] \right\}\;,\label{eq:a1ph1'}
 \eeq
\beq
{\cal M}^{H}(B_c \to a_1^+  h^{''})&=&  V_{cb}^* V_{ud} \left\{ \frac{- \sin\theta_{1}}{\sqrt{3}}[f_{B_c} (F_{fa;H}^{a_1^+
h_{1}^u}+F_{fa;H}^{h_{1}^d a_1^+})a_1
\right. \non && \left.
+ (M_{na;H}^{a_1^+ h_{1}^u}+M_{na;H}^{h_{1}^d a_1^+})  C_1 ] +\frac{\cos\theta_{1}}{\sqrt{6}}[f_{B_c}
\right. \non && \left.
\cdot (F_{fa;H}^{a_1^+ h_{8}^u}+F_{fa;H}^{h_{8}^d a_1^+}) a_1
 + (M_{na;H}^{a_1^+ h_{8}^u}+M_{na;H}^{h_{8}^d a_1^+ }) C_1 ] \right\}\;,\label{eq:a1ph1''}
\eeq
\beq
{\cal M}^{H}(B_c \to b_1^+  h')&=&  V_{cb}^* V_{ud} \left\{ \frac{\cos\theta_{1}}{\sqrt{3}}[f_{B_c} (F_{fa;H}^{b_1^+
h_{1}^u}+F_{fa;H}^{h_{1}^d b_1^+})a_1
\right. \non && \left.
+ (M_{na;H}^{b_1^+ h_{1}^u}+M_{na;H}^{h_{1}^d b_1^+}) C_1 ] +\frac{\sin\theta_{1}}{\sqrt{6}}[f_{B_c}
 \right. \non && \left.
\cdot (F_{fa;H}^{b_1^+ h_{8}^u}+F_{fa;H}^{h_{8}^d b_1^+}) a_1
 + (M_{na;H}^{b_1^+ h_{8}^u}+M_{na;H}^{h_{8}^d b_1^+ }) C_1 ] \right\}\;,\label{eq:b1ph1'}
 \eeq
\beq
{\cal M}^{H}(B_c \to b_1^+  h^{''})&=&  V_{cb}^* V_{ud} \left\{\frac{-\sin\theta_{1}}{\sqrt{3}}[f_{B_c} (F_{fa;H}^{b_1^+
h_{1}^u}+F_{fa;H}^{h_{1}^d b_1^+})a_1
\right. \non && \left.
+ (M_{na;H}^{b_1^+ h_{1}^u}+M_{na;H}^{h_{1}^d b_1^+}) C_1 ] + \frac{\cos\theta_{1}}{\sqrt{6}}[f_{B_c}
\right. \non && \left.
\cdot (F_{fa;H}^{b_1^+ h_{8}^u}+F_{fa;H}^{h_{8}^d b_1^+}) a_1
 + (M_{na;H}^{b_1^+ h_{8}^u}+M_{na;H}^{h_{8}^d b_1^+ }) C_1 ] \right\}\;,\label{eq:b1ph1''}
\eeq
\beq
{\cal M}^{H}(B_c \to \ov{K'}^{0} {K'}^+) &=& V_{cb}^* V_{ud}\left\{-\sin^2\theta_{K} (f_{B_c} F_{fa;H}^{\ov{K_{1A}}^0 K_{1A}}a_1 +
M_{na;H}^{\ov{K_{1A}}^0 K_{1A}} C_1)
\right. \non && \left.
-  \cos\theta_K \sin\theta_{K}(f_{B_c} F_{fa;H}^{\ov{K_{1A}}^0 K_{1B}}a_1+M_{na;H}^{\ov{K_{1A}}^0 K_{1B}} C_1 )
\right. \non && \left.
+ \cos\theta_{K} \sin\theta_{K}(f_{B_c} F_{fa;H}^{\ov{K_{1B}}^0 K_{1A}}a_1 +
M_{na;H}^{\ov{K_{1B}}^0 K_{1A}} C_1)
\right. \non && \left.
+ \cos^2\theta_{K} (f_{B_c} F_{fa;H}^{\ov{K_{1B}}^0 K_{1B}}a_1 +
M_{na;H}^{\ov{K_{1B}}^0 K_{1B}} C_1) \right\}\;,\label{eq:k127k127}
\eeq
\beq
{\cal M}^{H}(B_c \to \ov{K'}^{0} {K^{''}}^+) &=& V_{cb}^* V_{ud}\left\{-\cos\theta_{K}\sin\theta_{K} (f_{B_c} F_{fa;H}^{\ov{K_{1A}}^0 K_{1A}}a_1 +
M_{na;H}^{\ov{K_{1A}}^0 K_{1A}} C_1)
\right. \non && \left.
+  \sin^2\theta_{K}(f_{B_c} F_{fa;H}^{\ov{K_{1A}}^0 K_{1B}}a_1
+M_{na;H}^{\ov{K_{1A}}^0 K_{1B}} C_1 )
\right. \non && \left.
+ \cos^2\theta_{K}(f_{B_c} F_{fa;H}^{\ov{K_{1B}}^0 K_{1A}}a_1 +
M_{na;H}^{\ov{K_{1B}}^0 K_{1A}} C_1)
\right. \non && \left.
-\cos\theta_{K} \sin\theta_{K} (f_{B_c} F_{fa;H}^{\ov{K_{1B}}^0 K_{1B}}a_1 +
M_{na;H}^{\ov{K_{1B}}^0 K_{1B}} C_1) \right\}\;,\label{eq:k127k140}
\eeq
\beq
{\cal M}^{H}(B_c \to \ov{K^{''}}^0 {K'}^+) &=& V_{cb}^* V_{ud}\left\{\cos\theta_{K} \sin\theta_{K} (f_{B_c} F_{fa;H}^{\ov{K_{1A}}^0 K_{1A}}a_1 +
M_{na;H}^{\ov{K_{1A}}^0 K_{1A}} C_1)
\right. \non && \left.
+ \cos^2\theta_{K}(f_{B_c} F_{fa;H}^{\ov{K_{1A}}^0 K_{1B}}a_1
+M_{na;H}^{\ov{K_{1A}}^0 K_{1B}} C_1 )
\right. \non && \left.
+ \sin^2\theta_{K}(f_{B_c} F_{fa;H}^{\ov{K_{1B}}^0 K_{1A}}a_1 +
M_{na;H}^{\ov{K_{1B}}^0 K_{1A}} C_1)
\right. \non && \left.
+  \cos\theta_{K}\sin\theta_{K} (f_{B_c} F_{fa;H}^{\ov{K_{1B}}^0 K_{1B}}a_1 +
M_{na;H}^{\ov{K_{1B}}^0 K_{1B}} C_1) \right\}\;,\label{eq:k140k127}
\eeq
\beq
{\cal M}^{H}(B_c \to \ov{K^{''}}^0 {K^{''}}^+) &=& V_{cb}^* V_{ud}\left\{\cos^2\theta_{K} (f_{B_c} F_{fa;H}^{\ov{K_{1A}}^0 K_{1A}}a_1 +
M_{na;H}^{\ov{K_{1A}}^0 K_{1A}} C_1)
\right. \non && \left.
- \cos\theta_{K} \sin\theta_{K}(f_{B_c} F_{fa;H}^{\ov{K_{1A}}^0 K_{1B}}a_1
+M_{na;H}^{\ov{K_{1A}}^0 K_{1B}} C_1 )
\right. \non && \left.
 + \cos\theta_{K} \sin\theta_{K}(f_{B_c} F_{fa;H}^{\ov{K_{1B}}^0 K_{1A}}a_1 +
M_{na;H}^{\ov{K_{1B}}^0 K_{1A}} C_1)
\right. \non && \left.
-\sin^2\theta_{K} (f_{B_c} F_{fa;H}^{\ov{K_{1B}}^0 K_{1B}}a_1 +
M_{na;H}^{\ov{K_{1B}}^0 K_{1B}} C_1) \right\}\;. \label{eq:k140k140}
\eeq


\item[]{(ii)} For $\Delta S=1$ processes,
\beq
{\cal M}^{H}(B_c \to  {K'}^0 a_1^+) &=& \sqrt{2}{\cal M}^{H}(B_c \to  {K'}^+ a_1^0)
\non
 &=&  V_{cb}^* V_{us} \left\{\sin\theta_{K} [f_{B_c}
F_{fa;H}^{K_{1A}^0 a_1^+}a_1 + M_{na;H}^{K_{1A}^0 a_1^+} C_1]
\right. \non && \left.
+ \cos\theta_{K} [f_{B_c} F_{fa;H}^{K_{1B}^0 a_1^+}a_1 + M_{na;H}^{K_{1B}^0 a_1^+} C_1] \right\} \;,\label{eq:k127a1}\\
{\cal M}^{H}(B_c \to  {K^{''}}^0 a_1^+) &=& \sqrt{2}{\cal M}^{H}(B_c \to  {K^{''}}^+ a_1^0)
\non
 &=&  V_{cb}^* V_{us} \left\{\cos\theta_{K} [f_{B_c}
F_{fa;H}^{K_{1A}^0 a_1^+}a_1 + M_{na;H}^{K_{1A}^0 a_1^+} C_1]
\right. \non && \left.
- \sin\theta_{K} [f_{B_c} F_{fa;H}^{K_{1B}^0 a_1^+}a_1 + M_{na;H}^{K_{1B}^0 a_1^+} C_1] \right\} \;,\label{eq:k140a1}
\eeq
\beq
{\cal M}^{H}(B_c \to  {K'}^0 b_1^+) &=& \sqrt{2}{\cal M}^{H}(B_c \to  {K'}^+ b_1^0)
\non
 &=&  V_{cb}^* V_{us} \left\{\sin\theta_{K} [f_{B_c}
F_{fa;H}^{K_{1A}^0 b_1^+}a_1 + M_{na;H}^{K_{1A}^0 b_1^+} C_1]
\right. \non && \left.
+ \cos\theta_{K} [f_{B_c} F_{fa;H}^{K_{1B}^0 b_1^+}a_1 + M_{na;H}^{K_{1B}^0 b_1^+} C_1] \right\} \;,\label{eq:k127b1}
\eeq
\beq
{\cal M}^{H}(B_c \to  {K^{''}}^0 b_1^+) &=& \sqrt{2}{\cal M}^{H}(B_c \to  {K^{''}}^+ b_1^0)
\non
 &=&  V_{cb}^* V_{us} \left\{\cos\theta_{K} [f_{B_c}
F_{fa;H}^{K_{1A}^0 b_1^+}a_1 + M_{na;H}^{K_{1A}^0 b_1^+} C_1]
\right. \non && \left.
- \sin\theta_{K} [f_{B_c} F_{fa;H}^{K_{1B}^0 b_1^+}a_1 + M_{na;H}^{K_{1B}^0 b_1^+} C_1] \right\} \;,\label{eq:k140b1}
\eeq
\beq
{\cal M}^{H}(B_c \to {K'}^+  f')&=&  V_{cb}^* V_{us} \left\{ \frac{\cos\theta_{3}\sin\theta_{K}}{\sqrt{3}}[f_{B_c} (F_{fa;H}^{K_{1A}
f_{1}^u}+F_{fa;H}^{f_{1}^s K_{1A}})a_1
\right. \non && \left.
+ (M_{na;H}^{K_{1A} f_{1}^u}+M_{na;H}^{f_{1}^s K_{1A}}) C_1 ]+ \frac{\sin\theta_{3}\sin\theta_{K}}{\sqrt{6}}[f_{B_c}
\right. \non && \left.
\cdot (F_{fa;H}^{K_{1A} f_{8}^u}-2 F_{fa;H}^{f_{8}^s K_{1A}}) a_1 + (M_{na;H}^{K_{1A} f_{8}^u}-2 M_{na;H}^{f_{8}^s K_{1A} })
\right. \non && \left.
\cdot C_1 ]+ \frac{\cos\theta_{3}\cos\theta_{K}}{\sqrt{3}}[f_{B_c} (F_{fa;H}^{K_{1B} f_{1}^u}+F_{fa;H}^{f_{1}^s K_{1B}})a_1
\right. \non && \left.
+ (M_{na;H}^{K_{1B} f_{1}^u}+M_{na;H}^{f_{1}^s K_{1B}}) C_1 + \frac{\cos\theta_{K}\sin\theta_{3}}{\sqrt{6}}[f_{B_c}
\right. \non && \left.
\cdot (F_{fa;H}^{K_{1B} f_{8}^u}-2 F_{fa;H}^{f_{8}^s K_{1B}}) a_1 + (M_{na;H}^{K_{1B} f_{8}^u}
-2 M_{na;H}^{f_{8}^s K_{1B} }) C_1 ] \right\}\;,\label{eq:k127f'}
\eeq
\beq
{\cal M}^{H}(B_c \to {K'}^+  f^{''})&=&  V_{cb}^* V_{us} \left\{ \frac{- \sin\theta_{3}\sin\theta_{K}}{\sqrt{3}}[f_{B_c} (F_{fa;H}^{K_{1A}
f_{1}^u}+F_{fa;H}^{f_{1}^s K_{1A}})a_1
\right. \non && \left.
+ (M_{na;H}^{K_{1A} f_{1}^u}+M_{na;H}^{f_{1}^s K_{1A}}) C_1 ] + \frac{\cos\theta_{3}\sin\theta_{K}}{\sqrt{6}}[f_{B_c}
\right. \non && \left.
\cdot (F_{fa;H}^{K_{1A} f_{8}^u}-2 F_{fa;H}^{f_{8}^s K_{1A}}) a_1 + (M_{na;H}^{K_{1A} f_{8}^u}-2 M_{na;H}^{f_{8}^s K_{1A} })
\right. \non && \left.
\cdot C_1 ]-\frac{\cos\theta_{K} \sin\theta_{3}}{\sqrt{3}}[f_{B_c} (F_{fa;H}^{K_{1B} f_{1}^u}+F_{fa;H}^{f_{1}^s K_{1B}})a_1
\right. \non && \left.
+ (M_{na;H}^{K_{1B} f_{1}^u}+M_{na;H}^{f_{1}^s K_{1B}}) C_1 ] + \frac{\cos\theta_{K}\cos\theta_{3}}{\sqrt{6}}[f_{B_c}
\right. \non && \left.
\cdot (F_{fa;H}^{K_{1B} f_{8}^u}-2 F_{fa;H}^{f_{8}^s K_{1B}}) a_1 + (M_{na;H}^{K_{1B} f_{8}^u}
-2 M_{na;H}^{f_{8}^s K_{1B} }) C_1 ] \right\}\;,\label{eq:k127f''}
\eeq
\beq
{\cal M}^{H}(B_c \to {K^{''}}^+  f')&=&  V_{cb}^* V_{us} \left\{ \frac{\cos\theta_{3}\cos\theta_{K}}{\sqrt{3}}[f_{B_c} (F_{fa;H}^{K_{1A}
f_{1}^u}+F_{fa;H}^{f_{1}^s K_{1A}})a_1
\right. \non && \left.
+ (M_{na;H}^{K_{1A} f_{1}^u}+M_{na;H}^{f_{1}^s K_{1A}}) C_1 ]
 + \frac{\cos\theta_{K} \sin\theta_{3}}{\sqrt{6}}[f_{B_c}
\right. \non && \left.
\cdot (F_{fa;H}^{K_{1A} f_{8}^u}-2 F_{fa;H}^{f_{8}^s K_{1A}}) a_1 + (M_{na;H}^{K_{1A} f_{8}^u}-2 M_{na;H}^{f_{8}^s K_{1A} })
\right. \non && \left.
\cdot C_1 ]- \frac{\cos\theta_{3}\sin\theta_{K}}{\sqrt{3}}[f_{B_c} (F_{fa;H}^{K_{1B}
f_{1}^u}+F_{fa;H}^{f_{1}^s K_{1B}})a_1
\right. \non && \left.
+ (M_{na;H}^{K_{1B} f_{1}^u}+M_{na;H}^{f_{1}^s K_{1B}}) C_1 ]
 - \frac{\sin\theta_{K}\sin\theta_{3}}{\sqrt{6}}[f_{B_c}
\right. \non && \left.
\cdot (F_{fa;H}^{K_{1B} f_{8}^u}-2 F_{fa;H}^{f_{8}^s K_{1B}}) a_1 + (M_{na;H}^{K_{1B} f_{8}^u}
-2 M_{na;H}^{f_{8}^s K_{1B} }) C_1 ] \right\}\;,\label{eq:k140f'}
\eeq
\beq
{\cal M}^{H}(B_c \to {K^{''}}^+  f^{''})&=&  V_{cb}^* V_{us} \left\{\frac{-\cos\theta_{K} \sin\theta_{3}}{\sqrt{3}}[f_{B_c} (F_{fa;H}^{K_{1A}
f_{1}^u}+F_{fa;H}^{f_{1}^s K_{1A}})a_1
\right. \non && \left.
+ (M_{na;H}^{K_{1A} f_{1}^u}+M_{na;H}^{f_{1}^s K_{1A}}) C_1 ] + \frac{\cos\theta_{3}\cos\theta_{K}}{\sqrt{6}}[f_{B_c}
\right. \non && \left.
\cdot (F_{fa;H}^{K_{1A} f_{8}^u}-2 F_{fa;H}^{f_{8}^s K_{1A}}) a_1 + (M_{na;H}^{K_{1A} f_{8}^u}-2 M_{na;H}^{f_{8}^s K_{1A} })
\right. \non && \left.
\cdot C_1 ]+ \frac{\sin\theta_{3}\sin\theta_{K}}{\sqrt{3}}[f_{B_c} (F_{fa;H}^{K_{1B} f_{1}^u}+F_{fa;H}^{f_{1}^s K_{1B}})a_1
\right. \non && \left.
+ (M_{na;H}^{K_{1B} f_{1}^u}+M_{na;H}^{f_{1}^s K_{1B}}) C_1 ] - \frac{\cos\theta_{3}\sin\theta_{K}}{\sqrt{6}}[f_{B_c}
\right. \non && \left.
\cdot (F_{fa;H}^{K_{1B} f_{8}^u}-2 F_{fa;H}^{f_{8}^s K_{1B}}) a_1 + (M_{na;H}^{K_{1B} f_{8}^u}
-2 M_{na;H}^{f_{8}^s K_{1B} }) C_1 ] \right\}\;,\label{eq:k140f''}
\eeq
\beq
{\cal M}^{H}(B_c \to {K'}^+  h')&=&  V_{cb}^* V_{us} \left\{ \frac{\cos\theta_{3}\sin\theta_{K}}{\sqrt{3}}[f_{B_c} (F_{fa;H}^{K_{1A}
h_{1}^u}+F_{fa;H}^{h_{1}^s K_{1A}})a_1
\right. \non && \left.
+ (M_{na;H}^{K_{1A} h_{1}^u}+M_{na;H}^{h_{1}^s K_{1A}}) C_1 ] + \frac{\sin\theta_{3}\sin\theta_{K}}{\sqrt{6}}[f_{B_c}
\right. \non && \left.
\cdot (F_{fa;H}^{K_{1A} h_{8}^u}-2 F_{fa;H}^{h_{8}^s K_{1A}}) a_1 + (M_{na;H}^{K_{1A} h_{8}^u}-2 M_{na;H}^{h_{8}^s K_{1A} })
\right. \non && \left.
\cdot C_1 ]+ \frac{\cos\theta_{3}\cos\theta_{K}}{\sqrt{3}}[f_{B_c} (F_{fa;H}^{K_{1B} h_{1}^u}+F_{fa;H}^{h_{1}^s K_{1B}})a_1
\right. \non && \left.
+ (M_{na;H}^{K_{1B} h_{1}^u}+M_{na;H}^{h_{1}^s K_{1B}}) C_1 ] + \frac{\cos\theta_{K}\sin\theta_{3}}{\sqrt{6}}[f_{B_c}
\right. \non && \left.
\cdot (F_{fa;H}^{K_{1B} h_{8}^u}-2 F_{fa;H}^{h_{8}^s K_{1B}}) a_1 + (M_{na;H}^{K_{1b} h_{8}^u}
-2 M_{na;H}^{h_{8}^s K_{1B} }) C_1 ] \right\}\;,\label{eq:k127h'}
\eeq
\beq
{\cal M}^{H}(B_c \to {K'}^+  h^{''})&=&  V_{cb}^* V_{us} \left\{\frac{- \sin\theta_{3}\sin\theta_{K}}{\sqrt{3}}[f_{B_c} (F_{fa;H}^{K_{1A}
h_{1}^u}+F_{fa;H}^{h_{1}^s K_{1A}})a_1
 \right. \non && \left.
 + (M_{na;H}^{K_{1A} h_{1}^u}+M_{na;H}^{h_{1}^s K_{1A}}) C_1 ] + \frac{\cos\theta_{3}\sin\theta_{K}}{\sqrt{6}}[f_{B_c}
\right. \non && \left.
\cdot (F_{fa;H}^{K_{1A} h_{8}^u}-2 F_{fa;H}^{h_{8}^s K_{1A}}) a_1 + (M_{na;H}^{K_{1A} h_{8}^u}-2 M_{na;H}^{h_{8}^s K_{1A} })
\right. \non && \left.
\cdot C_1 ]-\frac{\cos\theta_{K} \sin\theta_{3}}{\sqrt{3}}[f_{B_c} (F_{fa;H}^{K_{1B} h_{1}^u}+F_{fa;H}^{h_{1}^s K_{1B}})a_1
\right. \non && \left.
+ (M_{na;H}^{K_{1B} h_{1}^u}+M_{na;H}^{h_{1}^s K_{1B}}) C_1 ] + \frac{\cos\theta_{K}\cos\theta_{3}}{\sqrt{6}}[f_{B_c}
\right. \non && \left.
\cdot (F_{fa;H}^{K_{1B} h_{8}^u}-2 F_{fa;H}^{h_{8}^s K_{1B}}) a_1 + (M_{na;H}^{K_{1b} h_{8}^u}
-2 M_{na;H}^{h_{8}^s K_{1B} }) C_1 ] \right\}\;,\label{eq:k127h''}
\eeq
\beq
{\cal M}^{H}(B_c \to {K^{''}}^+  h')&=&  V_{cb}^* V_{us} \left\{ \frac{\cos\theta_{3}\cos\theta_{K}}{\sqrt{3}}[f_{B_c} (F_{fa;H}^{K_{1A}
h_{1}^u}+F_{fa;H}^{h_{1}^s K_{1A}})a_1
\right. \non && \left.
+ (M_{na;H}^{K_{1A} h_{1}^u}+M_{na;H}^{h_{1}^s K_{1A}}) C_1 ] + \frac{\cos\theta_{K}\sin\theta_{3}}{\sqrt{6}}[f_{B_c}
\right. \non && \left.
\cdot (F_{fa;H}^{K_{1A} h_{8}^u}-2 F_{fa;H}^{h_{8}^s K_{1A}}) a_1 + (M_{na;H}^{K_{1A} h_{8}^u}-2 M_{na;H}^{h_{8}^s K_{1A} })
\right. \non && \left.
\cdot C_1 ]- \frac{\cos\theta_{3}\sin\theta_{K}}{\sqrt{3}}[f_{B_c} (F_{fa;H}^{K_{1B} h_{1}^u}+F_{fa;H}^{h_{1}^s K_{1B}})a_1
\right. \non && \left.
+ (M_{na;H}^{K_{1B} h_{1}^u}+M_{na;H}^{h_{1}^s K_{1B}}) C_1 ] - \frac{\sin\theta_{K}\sin\theta_{3}}{\sqrt{6}}[f_{B_c}
\right. \non && \left.
\cdot (F_{fa;H}^{K_{1B} h_{8}^u}-2 F_{fa;H}^{h_{8}^s K_{1B}}) a_1 + (M_{na;H}^{K_{1b} h_{8}^u}
-2 M_{na;H}^{h_{8}^s K_{1B} }) C_1 ] \right\}\;,\label{eq:k140h'}
\eeq
\beq
{\cal M}^{H}(B_c \to {K^{''}}^+  h^{''})&=&  V_{cb}^* V_{us} \left\{\frac{-\cos\theta_{K} \sin\theta_{3}}{\sqrt{3}}[f_{B_c} (F_{fa;H}^{K_{1A}
h_{1}^u}+F_{fa;H}^{h_{1}^s K_{1A}})a_1
\right. \non && \left.
+ (M_{na;H}^{K_{1A} h_{1}^u}+M_{na;H}^{h_{1}^s K_{1A}}) C_1 ] + \frac{\cos\theta_{3}\cos\theta_{K}}{\sqrt{6}}[f_{B_c}
\right. \non && \left.
\cdot (F_{fa;H}^{K_{1A} h_{8}^u}-2 F_{fa;H}^{h_{8}^s K_{1A}}) a_1 + (M_{na;H}^{K_{1A} h_{8}^u}-2 M_{na;H}^{h_{8}^s K_{1A} })
\right. \non && \left.
\cdot C_1 ]+ \frac{\sin\theta_{3}\sin\theta_{K}}{\sqrt{3}}[f_{B_c} (F_{fa;H}^{K_{1B} h_{1}^u}+F_{fa;H}^{h_{1}^s K_{1B}})a_1
 \right. \non && \left.
 + (M_{na;H}^{K_{1B} h_{1}^u}+M_{na;H}^{h_{1}^s K_{1B}}) C_1 ] -\frac{ \cos\theta_{3}\sin\theta_{K}}{\sqrt{6}}[f_{B_c}
\right. \non && \left.
\cdot (F_{fa;H}^{K_{1B} h_{8}^u}-2 F_{fa;H}^{h_{8}^s K_{1B}}) a_1 + (M_{na;H}^{K_{1b} h_{8}^u}
-2 M_{na;H}^{h_{8}^s K_{1B} }) C_1 ] \right\}\;.\label{eq:k140h''}
\eeq
\end{itemize}

\section{Numerical Results and Discussions}\label{sec:3}

In this section, we will calculate numerically the BRs and
polarization fractions for those considered sixty two $B_c \to VA/AV, AA$
decay modes. First of all, the central values of the input
parameters to be used are given in the following,
\begin{itemize}
\item[]{Masses (GeV):}
\beq
 m_W &=& 80.41;  \quad m_{B_c} = 6.286;  \quad m_b = 4.8;\;\; \quad m_c = 1.5; \non
m_\phi &=& 1.02;\; \quad  m_{K^*}= 0.892; \quad  m_{\rho} = 0.770; \;\;  m_{\omega}=0.782;   \non
 m_{a_1} &=& 1.23;  \quad  m_{K_{1A}} = 1.32; \quad   m_{f_1}= 1.28;  \quad  m_{f_8} =1.28;    \non
m_{b_1}&=& 1.21;  \quad  m_{K_{1B}} =1.34; \quad  m_{h_1}=1.23;  \quad  m_{h_8} = 1.37; \label{eq:mass}
\eeq
\item[]{Decay constants (GeV):}
\beq
f_{\phi} &=& 0.231;\quad  f_{\phi}^T = 0.200;  \quad f_{K^*} = 0.217; \quad f_{K^*}^T = 0.185; \non
f_{\rho}&=& 0.209; \quad f^T_{\rho}= 0.165;\;\;\; \quad f_{\omega}= 0.195; \quad f_{\omega}^T = 0.145; \non
f_{a_1} &=& 0.238; \quad f_{K_{1A}} = 0.250; \quad f_{f_1}= 0.245; \quad f_{f_8}= 0.239;  \non
f_{b_1} &=& 0.180; \quad f_{K_{1B}}= 0.190;  \quad f_{h_1}= 0.180; \quad f_{h_8}= 0.190; \non
 f_{B_c} &=& 0.489; \label{eq:dconst}
 \eeq
\item[]{ QCD scale and $B_c$ meson lifetime:}
\beq
\Lambda_{\overline{\mathrm{MS}}}^{(f=4)} &=& 0.250\; {\rm GeV},  \quad
\tau_{B_c^+}= 0.46\; {\rm ps}.
\eeq

\end{itemize}

For the CKM matrix elements, here we adopt the Wolfenstein
parametrization for the CKM matrix, and take $A=0.814$ and
 $\lambda=0.2257$, $\bar{\rho}=0.135$ and $\bar{\eta}=0.349$ \cite{Amsler08:pdg}.
 In numerical calculations, central values of input parameters will be
used implicitly unless otherwise stated.

For these considered $B_c \to M_2 M_3$ decays, the decay rate can be written explicitly as,
\beq
\Gamma =\frac{G_{F}^{2}|\bf{P_c}|}{16 \pi m^{2}_{B_c} }
\sum_{\sigma=L,T}{\cal M}^{(\sigma)\dagger }{\cal M^{(\sigma)}}\;
\label{dr1}
\eeq
where $|\bf{P_c}|\equiv |\bf{P_{2z}}|=|\bf{P_{3z}}|$ is the momentum of either of the
outgoing vector or axial-vector mesons.

Based on the helicity amplitudes~(\ref{eq:amp}), we can define the
transversity amplitudes,
\beq
{\cal A}_{L}&=&-\xi m^{2}_{B_c}{\cal M}_{L}, \quad {\cal A}_{\parallel}=\xi
\sqrt{2}m^{2}_{B_c}{\cal M}_{N}, \quad {\cal A}_{\perp}=\xi
m^{2}_{B_c} \sqrt{2(r^{2}-1)} {\cal M }_{T}\;. \label{eq:ase}
\eeq
for the longitudinal, parallel, and perpendicular polarizations,
respectively, with the normalization factor
$\xi=\sqrt{G^2_{F}{\bf{P_c}} /(16\pi m^2_{B_c}\Gamma)}$ and the
ratio $r=P_{2}\cdot P_{3}/(m_2\; m_3)$.
These amplitudes satisfy the relation,
\beq
|{\cal A}_{L}|^2+|{\cal A}_{\parallel}|^2+|{\cal A}_{\perp}|^2=1
\eeq
following the summation in Eq.~(\ref{dr1}).

The polarization fractions $f_{L},f_{||}$ and $f_{\perp}$ can be defined as,
\beq
f_{L(||,\perp)}= \frac{|{\cal A}_{L(||,\perp)}|^2}{|{\cal
A}_L|^2+|{\cal A}_{||}|^2+|{\cal A}_{\perp}|^2},
\label{eq:pf}
\eeq

With the analytic formulas for the complete decay amplitudes as given explicitly in Eqs.~(\ref{eq:rhopa1}-\ref{eq:k140h''}), by employing
the input parameters 
and Eq.(\ref{dr1}), we calculate and then present the pQCD predictions for the
{\it CP}-averaged BRs and longitudinal polarization fractions (LPFs)
of the considered decays with errors in Tables~\ref{tab:bra1-b1v}-\ref{tab:brk1a-k1bha}.
The dominant errors come from the uncertainties of charm quark mass
$m_c=1.5 \pm 0.15$ GeV and the combined Gegenbauer moments $a_i$ of the relevant
meson distribution amplitudes, respectively.

\begin{table}[b]
\caption{The pQCD predictions of BRs and LPFs for
$B_c \to (a_1, b_1) (\rho, K^*, \omega)$ decays. The source of the dominant errors
is explained in the text.} \label{tab:bra1-b1v}
\begin{center}\vspace{-0.5cm}
\begin{tabular}[t]{l|lc|l|lc} \hline  \hline
 $\Delta S =0 $  &  \multicolumn{2}{|c|} {}                & $\Delta S =0 $  & \multicolumn{2}{|c}{}  \\
  Decay modes    &  BRs $(10^{-7})$ & LPFs $(\%)$        & Decay modes     & BRs $(10^{-6})$ & LPFs $(\%)$  \\
\hline
  $\rm{B_c \to \rho^+ a_1^0}$ &$7.5^{+0.1}_{-0.2}(m_c)^{+3.0}_{-2.8}(a_i)$& $99.2^{+0.3}_{-0.4}$
 &$\rm{B_c \to \rho^+ b_1^0}$ &$9.9^{+4.7}_{-3.9}(m_c)^{+5.6}_{-4.4}(a_i)$& $93.8^{+2.5}_{-4.4}$  \\
  $\rm{B_c \to a_1^+ \rho^0}$ &$7.5^{+0.1}_{-0.2}(m_c)^{+3.0}_{-2.8}(a_i)$& $99.2^{+0.3}_{-0.4}$
 &$\rm{B_c \to b_1^+ \rho^0}$ &$9.9^{+4.5}_{-4.0}(m_c)^{+5.7}_{-4.2}(a_i)$& $93.8^{+2.5}_{-4.4}$  \\
 \hline
  $\rm{B_c \to a_1^+ \omega}$ &$20.2^{+3.0}_{-0.0}(m_c)^{+7.8}_{-4.8}(a_i)$&$84.7^{+5.0}_{-4.4}$
 &$\rm{B_c \to b_1^+ \omega}$ &$0.6^{+0.5}_{-0.4}(m_c)^{+0.3}_{-0.4}(a_i)$& $96.3^{+2.3}_{-7.5}$  \\
 \hline    \hline
$\Delta S =1 $  &  \multicolumn{2}{|c|} {}                   & $\Delta S =1 $  & \multicolumn{2}{|c}{}  \\
  Decay modes   &  BRs $(10^{-8})$ & LPFs $(\%)$            & Decay modes     &   BRs $(10^{-7})$ & LPFs $(\%)$  \\
\hline
  $\rm{B_c \to a_1^+ {K^*}^{0}}$ &$6.5^{+0.2}_{-0.0}(m_c)^{+3.5}_{-2.5}(a_i)$& $83.6^{+5.3}_{-7.5}$
 &$\rm{B_c \to b_1^+ {K^*}^{0}}$ &$3.6^{+1.1}_{-0.8}(m_c)^{+1.8}_{-1.6}(a_i)$& $91.2^{+2.5}_{-4.1}$ \\
  $\rm{B_c \to {K^*}^+ a_1^0}$ &$3.3^{+0.1}_{-0.0}(m_c)^{+1.7}_{-1.3}(a_i)$& $83.6^{+5.3}_{-7.5}$
 &$\rm{B_c \to {K^*}^+ b_1^0}$ &$1.8^{+0.5}_{-0.4}(m_c)^{+0.9}_{-0.7}(a_i)$& $91.2^{+2.5}_{-4.1}$  \\
 \hline    \hline
\end{tabular}
\end{center}
\end{table}


\begin{table}[t]
\caption{Same as Table~\ref{tab:bra1-b1v} but for
$B_c \to (a_1, b_1) (a_1, b_1)$ decays.} \label{tab:bra1-b1a}
\begin{center}\vspace{-0.6cm}
\begin{tabular}[t]{l|lc|l|lc} \hline  \hline
 $\Delta S =0 $  &  \multicolumn{2}{|c|} {}                & $\Delta S =0 $  & \multicolumn{2}{|c}{}  \\
  Decay modes    &  BRs $(10^{-5})$ & LPFs $(\%)$        & Decay modes     & BRs $(10^{-5})$ & LPFs $(\%)$  \\
\hline
 $\rm{B_c \to a_1^+ a_1^0}$  &$0.0$& --
 &$\rm{B_c \to b_1^+ b_1^0}$ &$0.0$& --  \\
  $\rm{B_c \to a_1^+ b_1^0}$ &$2.2^{+0.6}_{-0.5}(m_c)^{+1.1}_{-0.9}(a_i)$&  $92.4^{+1.9}_{-2.8}$
 &$\rm{B_c \to b_1^+ a_1^0}$ &$2.2^{+0.6}_{-0.5}(m_c)^{+1.1}_{-0.8}(a_i)$&  $91.8^{+2.0}_{-2.6}$\\
 \hline    \hline
\end{tabular}
\end{center}
\end{table}


\begin{table}[b]
\caption{Same as Table~\ref{tab:bra1-b1v} but for $B_c \to (\rho, K^*) (f_1(1285), f_1(1420))$ decays.}
\label{tab:brf1-f8v}
\begin{center}\vspace{-0.6cm}
\begin{tabular}[t]{l|lc|lc} \hline  \hline
$\Delta S =0$  &   \multicolumn{2}{|c|}{ $\theta_{3}=38^\circ$ }     &   \multicolumn{2}{|c}{ $\theta_{3}=50^\circ$} \\
 Decay modes       &    BRs $(10^{-6})$  &  LPFs $(\%)$                &    BRs$(10^{-6})$  &  LPFs $(\%)$  \\
\hline
 $\rm{B_c \to \rho^+ f_1(1285)}$                                              &$2.1^{+0.3}_{-0.0}(m_c)^{+0.8}_{-0.4}(a_i)$&     $82.6^{+5.4}_{-3.8}$
                                            &$1.9^{+0.3}_{-0.0}(m_c)^{+0.7}_{-0.3}(a_i)$&   $82.3^{+5.6}_{-3.8}$\\
 $\rm{B_c \to \rho^+ f_1(1420)}\times 10\footnotemark[1]$                 &$0.4^{+0.0}_{-0.1}(m_c)^{+0.8}_{-0.2}(a_i)$&     $88.7^{+11.6}_{-9.2}$
                                            &$2.2^{+0.2}_{-0.0}(m_c)^{+2.2}_{-0.8}(a_i)$&   $85.7^{+8.6}_{-6.9}$\\
 \hline    \hline
$\Delta S =1$ &   \multicolumn{2}{|c|}{ $\theta_{3}=38^\circ$ }     &   \multicolumn{2}{|c}{ $\theta_{3}=50^\circ$} \\
 Decay modes       &    BRs $(10^{-7})$  &  LPFs $(\%)$                &    BRs $(10^{-7})$  &  LPFs $(\%)$  \\
\hline
 $\rm{B_c \to {K^{*}}^+ f_1(1285)}\times 10$                                     &$1.6^{+0.3}_{-0.0}(m_c)^{+1.5}_{-0.6}(a_i)$&   $61.0^{+24.2}_{-22.2}$
                                            &$0.4^{+0.1}_{-0.0}(m_c)^{+0.5}_{-0.2}(a_i)$&  $33.7^{+56.0}_{-38.6}$\\
 $\rm{B_c \to {K^{*}}^+ f_1(1420)}$                                          &$1.1^{+0.1}_{-0.0}(m_c)^{+0.4}_{-0.4}(a_i)$&   $85.4^{+4.5}_{-5.8}$
                                            &$1.2^{+0.2}_{-0.0}(m_c)^{+0.5}_{-0.3}(a_i)$&  $83.9^{+5.2}_{-6.3}$\\
 \hline \hline
 \end{tabular}
  \footnotetext[1]{Here, the factor 10 is specifically used for the BRs. The following ones have the same meaning.}
\end{center}
\end{table}

\begin{table}[b]
\caption{Same as Table~\ref{tab:bra1-b1v} but for $B_c \to (\rho, K^*) (h_1(1170), h_1(1380))$ decays.}
\label{tab:brh1-h8v}
\begin{center}\vspace{-0.6cm}
\begin{tabular}[t]{l|lc|lc}
 \hline \hline
$\Delta S = 0 $        &   \multicolumn{2}{|c|}{ $\theta_{1}=10^\circ$ }     &   \multicolumn{2}{|c}{ $\theta_{1}=45^\circ$} \\
 Decay modes       &    BRs $(10^{-7})$  &  LPFs $(\%)$                &    BRs $(10^{-7})$  &  LPFs $(\%)$  \\
\hline
 $\rm{B_c \to \rho^+ h_1(1170)}$     &$6.4^{+4.6}_{-4.1}(m_c)^{+2.2}_{-3.9}(a_i)$&     $96.4^{+1.7}_{-7.0}$
                                                        &$8.6^{+7.4}_{-5.1}(m_c)^{+2.7}_{-3.4}(a_i)$&  $96.3^{+1.7}_{-5.4}$\\
 $\rm{B_c \to \rho^+ h_1(1380)}$ &$2.5^{+2.6}_{-0.8}(m_c)^{+2.0}_{-1.4}(a_i)$&     $96.3^{+2.1}_{-2.6}$
                                                        &$0.3^{+0.2}_{-0.2}(m_c)^{+0.6}_{-0.1}(a_i)$&  $98.0^{+1.7}_{-2.8}$\\
 \hline    \hline
$\Delta S = 1 $        &   \multicolumn{2}{|c|}{ $\theta_{1}=10^\circ$ }     &   \multicolumn{2}{|c}{ $\theta_{1}=45^\circ$} \\
 Decay modes       &    BRs $(10^{-7})$  &  LPFs $(\%)$                &    BRs $(10^{-7})$  &  LPFs $(\%)$  \\
\hline
 $\rm{B_c \to {K^{*}}^+ h_1(1170)}$     &$0.2^{+0.1}_{-0.1}(m_c)^{+0.0}_{-0.1}(a_i)$&    $82.3^{+6.2}_{-9.5}$
                                                       &$2.1^{+0.6}_{-0.6}(m_c)^{+1.1}_{-0.8}(a_i)$&  $89.5^{+3.4}_{-5.1}$\\
 $\rm{B_c \to {K^{*}}^+ h_1(1380)}$          &$4.8^{+2.2}_{-1.9}(m_c)^{+2.4}_{-2.0}(a_i)$&    $93.2^{+2.2}_{-4.5}$
                                                       &$2.9^{+1.7}_{-1.3}(m_c)^{+1.4}_{-1.1}(a_i)$&  $94.9^{+1.8}_{-4.1}$\\
\hline \hline
\end{tabular}
\end{center}
\end{table}

\begin{table}[b]
\caption{Same as Table~\ref{tab:bra1-b1v} but for $B_c \to  (a_1^+, b_1^+)(f_1(1285), f_1(1420))$ decays.}
\label{tab:brf1-f8a}
\begin{center}\vspace{-0.6cm}
\begin{tabular}[t]{l|lc|lc} \hline  \hline
$\Delta S =0$  &   \multicolumn{2}{|c|}{ $\theta_{3}=38^\circ$ }     &   \multicolumn{2}{|c}{ $\theta_{3}=50^\circ$} \\
 Decay modes       &    BRs $(10^{-6})$  &  LPFs $(\%)$                &    BRs $(10^{-6})$  &  LPFs $(\%)$  \\
\hline
 $\rm{B_c \to a_1(1260)^+ f_1(1285)}$              &$6.5^{+1.0}_{-0.9}(m_c)^{+0.5}_{-1.0}(a_i)$&  $83.6^{+2.4}_{-4.1}$
                                                            &$6.1^{+1.0}_{-0.9}(m_c)^{+0.4}_{-0.9}(a_i)$&  $84.0^{+2.3}_{-4.0}$\\
 $\rm{B_c \to a_1(1260)^+ f_1(1420)}\times 10$ &$0.3^{+0.1}_{-0.1}(m_c)^{+0.7}_{-0.3}(a_i)$&   $56.8^{+43.2}_{-56.8}$ 
                                                            &$3.9^{+0.7}_{-0.0}(m_c)^{+1.3}_{-1.6}(a_i)$&  $78.5^{+7.4}_{-13.9}$\\
 \hline    \hline
$\Delta S =0$ &   \multicolumn{2}{|c|}{ $\theta_{3}=38^\circ$ }     &   \multicolumn{2}{|c}{ $\theta_{K}=50^\circ$} \\
 Decay modes       &    BRs $(10^{-7})$  &  LPFs $(\%)$                &    BRs $(10^{-7})$  &  LPFs $(\%)$  \\
\hline
 $\rm{B_c \to b_1(1235)^+ f_1(1285)}$     &$2.8^{+4.1}_{-0.5}(m_c)^{+1.8}_{-0.9}(a_i)$&  $65.2^{+28.3}_{-16.4}$
                                                            &$3.0^{+4.4}_{-0.8}(m_c)^{+1.5}_{-0.9}(a_i)$&  $68.7^{+21.7}_{-14.6}$\\
 $\rm{B_c \to b_1(1235)^+ f_1(1420)}$ &$1.4^{+0.2}_{-0.1}(m_c)^{+0.7}_{-0.9}(a_i)$&  $100.0\pm 0.0$
                                                            &$1.2^{+0.4}_{-0.4}(m_c)^{+1.1}_{-0.8}(a_i)$&  $100.0^{+0.0}_{-0.8}$\\
 \hline \hline
 \end{tabular}
\end{center}
\end{table}

\begin{table}[t]
\caption{Same as Table~\ref{tab:bra1-b1v} but for $B_c \to  (a_1^+, b_1^+)(h_1(1170), h_1(1380))$ decays.}
\label{tab:brh1-h8a}
\begin{center}\vspace{-0.6cm}
\begin{tabular}[t]{l|lc|lc}
 \hline \hline
$\Delta S = 0 $        &   \multicolumn{2}{|c|}{ $\theta_{1}=10^\circ$ }     &   \multicolumn{2}{|c}{ $\theta_{1}=45^\circ$} \\
 Decay modes       &    BRs $(10^{-6})$  &  LPFs $(\%)$                &    BRs $(10^{-6})$  &  LPFs $(\%)$  \\
\hline
 $\rm{B_c \to a_1(1260)^+ h_1(1170)}$                     &$1.3^{+0.2}_{-0.5}(m_c)^{+0.5}_{-0.4}(a_i)$&  $86.3^{+2.0}_{-9.8}$
                                                            &$0.7^{+0.2}_{-0.4}(m_c)^{+0.3}_{-0.2}(a_i)$&  $73.1^{+7.4}_{-28.8}$\\
 $\rm{B_c \to a_1(1260)^+ h_1(1380)}\times 10$        &$1.1^{+0.7}_{-0.0}(m_c)^{+1.3}_{-0.5}(a_i)$&  $68.8^{+23.2}_{-11.5}$
                                                            &$6.8^{+0.2}_{-1.1}(m_c)^{+2.3}_{-2.4}(a_i)$&  $100.0^{+0.0}_{-1.2}$\\
 \hline    \hline
$\Delta S = 0 $        &   \multicolumn{2}{|c|}{ $\theta_{1}=10^\circ$ }     &   \multicolumn{2}{|c}{ $\theta_{1}=45^\circ$} \\
 Decay modes       &    BRs $(10^{-5})$  &  LPFs $(\%)$                &    BRs $(10^{-5})$  &  LPFs $(\%)$  \\
\hline
 $\rm{B_c \to b_1(1235)^+ h_1(1170)}$     &$8.1^{+3.6}_{-2.8}(m_c)^{+3.9}_{-3.4}(a_i)$&  $96.4^{+1.0}_{-1.6}$
                                                    &$10.3^{+4.0}_{-3.3}(m_c)^{+4.7}_{-3.8}(a_i)$&  $96.4^{+0.9}_{-1.4}$\\
 $\rm{B_c \to b_1(1235)^+ h_1(1380)}$ &$2.5^{+0.5}_{-0.7}(m_c)^{+1.4}_{-1.2}(a_i)$&  $100.0^{+0.0}_{-0.8}$
                                                    &$0.3^{+0.2}_{-0.1}(m_c)^{+0.5}_{-0.2}(a_i)$&  $100.0\pm 0.0$\\
\hline \hline
\end{tabular}
\end{center}
\end{table}

\begin{table}[b]
\caption{Same as Table~\ref{tab:bra1-b1v} but for $B_c \to (K_1(1270), K_1(1400)) (\rho, K^*, \omega, \phi)$ decays.}
\label{tab:brk1a-k1bv}
\begin{center}\vspace{-0.6cm}
\begin{tabular}[t]{l|lc|lc}
 \hline \hline
 $\Delta S =0$     &   \multicolumn{2}{|c|}{ $\theta_{K}=45^\circ$ }     &   \multicolumn{2}{|c}{ $\theta_{K}=-45^\circ$} \\
 Decay modes       &    BRs $(10^{-6})$  &  LPFs $(\%)$                &    BRs $(10^{-6})$  &  LPFs $(\%)$  \\
\hline
 $\rm{B_c \to \ov{K}^{*0} {K_1}(1270)^{+}}$ &$3.8^{+0.8}_{-0.8}(m_c)^{+3.1}_{-2.7}(a_i)$&  $96.8^{+2.5}_{-6.1}$
                                             &$2.3^{+0.5}_{-0.4}(m_c)^{+2.9}_{-1.3}(a_i)$&  $42.3^{+37.8}_{-30.2}$\\
 $\rm{B_c \to \ov{K}^{*0} {K_1(1400)}^+}$  &$2.2^{+0.5}_{-0.4}(m_c)^{+3.0}_{-1.2}(a_i)$& $42.7^{+37.9}_{-29.4}$
                                             &$3.8^{+0.8}_{-0.8}(m_c)^{+2.9}_{-2.8}(a_i)$&   $96.9^{+2.3}_{-6.1}$\\
  \hline
 $\rm{B_c \to \ov{K_1}(1270)^0 {K^{*}}^+}$     &$9.7^{+2.2}_{-2.5}(m_c)^{+5.3}_{-5.1}(a_i)$&  $97.5^{+2.6}_{-5.6}$
                                             &$5.6^{+2.4}_{-2.1}(m_c)^{+4.3}_{-3.3}(a_i)$&  $81.1^{+10.4}_{-17.8}$\\
 $\rm{B_c \to \ov{K_1}(1400)^0 {K^{*}}^+}$ &$5.8^{+2.2}_{-1.7}(m_c)^{+4.6}_{-3.1}(a_i)$&$82.4^{+9.3}_{-13.8}$
                                             &$9.6^{+2.2}_{-2.5}(m_c)^{+5.2}_{-5.1}(a_i)$&    $97.6^{+2.4}_{-5.7}$\\
  \hline    \hline
 $\Delta S =1$     &   \multicolumn{2}{|c|}{ $\theta_{K}=45^\circ$ }     &   \multicolumn{2}{|c}{ $\theta_{K}=-45^\circ$} \\
 Decay modes       &    BRs $(10^{-7})$  &  LPFs $(\%)$                &    BRs $(10^{-7})$  &  LPFs $(\%)$  \\
 \hline
 $\rm{B_c \to {K_1}(1270)^0 \rho^{+}}$     &$3.1^{+1.4}_{-1.1}(m_c)^{+2.6}_{-1.7}(a_i)$&  $89.5^{+5.9}_{-9.3}$
                                           &$4.0^{+1.0}_{-1.2}(m_c)^{+2.2}_{-2.1}(a_i)$&  $99.1^{+0.8}_{-3.6}$\\
 $\rm{B_c \to {K_1}(1400)^0 \rho^{+}}$ &$4.0^{+0.9}_{-1.2}(m_c)^{+2.1}_{-2.2}(a_i)$&  $99.1^{+0.8}_{-3.4}$
                                           &$3.0^{+1.4}_{-1.0}(m_c)^{+2.8}_{-1.5}(a_i)$&  $89.7^{+5.6}_{-9.3}$\\
\hline
 $\rm{B_c \to {K_1}(1270)^{+} \rho^0}$     &$1.5^{+0.7}_{-0.5}(m_c)^{+1.4}_{-0.7}(a_i)$&  $89.5^{+5.9}_{-9.3}$
                                           &$2.0^{+0.5}_{-0.6}(m_c)^{+1.1}_{-1.0}(a_i)$&  $99.1^{+0.8}_{-3.6}$\\
 $\rm{B_c \to {K_1}(1400)^{+} \rho^0}$ &$2.0^{+0.5}_{-0.6}(m_c)^{+1.1}_{-1.0}(a_i)$&  $99.1^{+0.8}_{-3.4}$
                                           &$1.5^{+0.7}_{-0.5}(m_c)^{+1.4}_{-0.8}(a_i)$&  $89.7^{+5.6}_{-9.3}$\\
\hline
 $\rm{B_c \to {K_1}(1270)^+ \omega}$      &$1.4^{+0.7}_{-0.5}(m_c)^{+1.2}_{-0.7}(a_i)$&  $89.8^{+5.4}_{-8.7}$
                                           &$1.7^{+0.5}_{-0.5}(m_c)^{+0.9}_{-0.8}(a_i)$&  $99.1^{+0.8}_{-3.9}$\\
 $\rm{B_c \to {K_1}(1400)^+ \omega}$  &$1.7^{+0.5}_{-0.5}(m_c)^{+0.9}_{-0.9}(a_i)$&  $99.1^{+0.8}_{-3.8}$
                                           &$1.4^{+0.6}_{-0.5}(m_c)^{+1.1}_{-0.7}(a_i)$&   $89.9^{+5.5}_{-8.8}$\\
\hline \hline
 $\rm{B_c \to {K_1}(1270)^+ \phi}$      &$1.9^{+0.2}_{-0.3}(m_c)^{+1.0}_{-1.4}(a_i)$&  $95.2^{+2.7}_{-10.9}$
                                           &$1.5^{+0.3}_{-0.4}(m_c)^{+1.2}_{-0.8}(a_i)$&  $29.9^{+30.8}_{-27.3}$\\
 $\rm{B_c \to {K_1}(1400)^+ \phi}$  &$1.4^{+0.4}_{-0.3}(m_c)^{+1.3}_{-0.6}(a_i)$&  $30.3^{+31.0}_{-27.6}$
                                           &$1.9^{+0.1}_{-0.3}(m_c)^{+1.0}_{-1.4}(a_i)$&  $95.2^{+2.9}_{-10.8}$\\
\hline \hline
\end{tabular}
\end{center}
\end{table}


\begin{table}[b]
\caption{Same as Table~\ref{tab:bra1-b1v} but for $B_c \to (K_1(1270), K_1(1400)) (a_1, b_1, K_1(1270), K_1(1400))$ decays.}
\label{tab:brk1a-k1ba}
\begin{center}\vspace{-0.6cm}
\begin{tabular}[t]{l|lc|lc}
 \hline \hline
 $\Delta S =0$     &   \multicolumn{2}{|c|}{ $\theta_{K}=45^\circ$ }     &   \multicolumn{2}{|c}{ $\theta_{K}=-45^\circ$} \\
 Decay modes       &    BRs $(10^{-5})$  &  LPFs $(\%)$                  &    BRs $(10^{-5})$  &  LPFs $(\%)$  \\
\hline
 $\rm{B_c \to \ov{K_1}(1270)^0 {K_1}(1270)^+}$      &$1.2^{+0.2}_{-0.1}(m_c)^{+1.8}_{-0.9}(a_i)$&  $99.7^{+0.1}_{-1.0}$
                                                              &$2.9^{+1.2}_{-1.0}(m_c)^{+4.4}_{-2.3}(a_i)$&  $71.9^{+16.2}_{-24.6}$\\
 $\rm{B_c \to \ov{K_1}(1270)^0 {K_1(1400)}^+}$     &$3.7^{+1.3}_{-1.1}(m_c)^{+3.1}_{-2.2}(a_i)$&  $96.2^{+3.5}_{-8.4}$
                                                              &$1.9^{+0.5}_{-0.5}(m_c)^{+2.2}_{-1.4}(a_i)$&  $94.8^{+3.4}_{-10.3}$\\
  \hline
 $\rm{B_c \to \ov{K_1}(1400)^0 {K_1}(1270)^+}$  &$1.9^{+0.5}_{-0.5}(m_c)^{+2.2}_{-1.4}(a_i)$&  $94.6^{+3.6}_{-10.7}$
                                                              &$3.7^{+1.3}_{-1.1}(m_c)^{+3.2}_{-2.1}(a_i)$&  $96.1^{+3.6}_{-8.6}$\\
 $\rm{B_c \to \ov{K_1}(1400)^0 {K_1}(1400)^+}$ &$2.8^{+1.2}_{-1.0}(m_c)^{+4.3}_{-2.3}(a_i)$&  $72.7^{+15.8}_{-24.3}$
                                                              &$1.1^{+0.2}_{-0.0}(m_c)^{+1.9}_{-0.9}(a_i)$&  $99.7^{+0.0}_{-1.0}$\\
  \hline    \hline
 $\Delta S =1$     &   \multicolumn{2}{|c|}{ $\theta_{K}=45^\circ$ }     &   \multicolumn{2}{|c}{ $\theta_{K}=-45^\circ$} \\
 Decay modes       &    BRs $(10^{-7})$  &  LPFs $(\%)$                  &    BRs $(10^{-7})$  &  LPFs $(\%)$  \\
 \hline
 $\rm{B_c \to {K_1}(1270)^0 a_1(1260)^+}$     &$4.6^{+1.3}_{-1.0}(m_c)^{+4.7}_{-2.4}(a_i)$&  $79.2^{+12.4}_{-16.3}$
                                                       &$8.3^{+1.3}_{-1.8}(m_c)^{+3.6}_{-3.9}(a_i)$&  $99.3^{+0.8}_{-5.5}$\\
 $\rm{B_c \to {K_1}(1400)^0 a_1(1260)^+}$ &$8.0^{+1.3}_{-1.7}(m_c)^{+3.5}_{-3.7}(a_i)$&  $100.0^{+0.0}_{-3.8}$
                                                       &$4.5^{+1.2}_{-1.1}(m_c)^{+4.4}_{-2.5}(a_i)$&  $81.3^{+12.5}_{-16.6}$\\
\hline
 $\rm{B_c \to {K_1}(1270)^+ a_1(1260)^0}$     &$2.3^{+0.6}_{-0.5}(m_c)^{+2.4}_{-1.3}(a_i)$&  $79.2^{+12.4}_{-16.3}$
                                                       &$4.2^{+0.6}_{-1.0}(m_c)^{+1.8}_{-2.0}(a_i)$&  $99.3^{+0.8}_{-5.5}$  \\
 $\rm{B_c \to {K_1}(1400)^+ a_1(1260)^0}$     &$4.0^{+0.7}_{-0.9}(m_c)^{+1.8}_{-1.9}(a_i)$&  $100.0^{+0.0}_{-3.8}$
                                                       &$2.2^{+0.6}_{-0.5}(m_c)^{+2.3}_{-1.1}(a_i)$&  $81.3^{+12.5}_{-16.6}$  \\
  \hline    \hline
 $\Delta S =1$     &   \multicolumn{2}{|c|}{ $\theta_{K}=45^\circ$ }     &   \multicolumn{2}{|c}{ $\theta_{K}=-45^\circ$} \\
 Decay modes       &    BRs $(10^{-6})$  &  LPFs $(\%)$                  &    BRs $(10^{-6})$  &  LPFs $(\%)$  \\
 \hline
 $\rm{B_c \to {K_1}(1270)^0 b_1(1235)^+}$     &$1.6^{+0.8}_{-0.5}(m_c)^{+1.3}_{-0.9}(a_i)$&  $91.3^{+5.0}_{-5.1}$
                                                       &$1.4^{+0.4}_{-0.2}(m_c)^{+0.8}_{-0.7}(a_i)$&  $100.0^{+0.0}_{-0.3}$\\
 $\rm{B_c \to {K_1}(1400)^0 b_1(1235)^+}$ &$1.3^{+0.4}_{-0.2}(m_c)^{+0.9}_{-0.5}(a_i)$&  $100.0 \pm 0.0$
                                                       &$1.5^{+0.8}_{-0.5}(m_c)^{+1.3}_{-0.9}(a_i)$&  $93.6^{+5.0}_{-5.1}$\\
\hline
 $\rm{B_c \to {K_1}(1270)^+ b_1(1235)^0}$     &$0.8^{+0.4}_{-0.3}(m_c)^{+0.6}_{-0.5}(a_i)$&  $91.4^{+4.9}_{-5.1}$
                                                       &$0.7^{+0.2}_{-0.1}(m_c)^{+0.4}_{-0.4}(a_i)$&  $100.0^{+0.0}_{-0.3}$  \\
 $\rm{B_c \to {K_1}(1400)^+ b_1(1235)^0}$     &$0.7^{+0.2}_{-0.2}(m_c)^{+0.4}_{-0.4}(a_i)$&  $100.0 \pm 0.0$
                                                       &$0.8^{+0.3}_{-0.3}(m_c)^{+0.6}_{-0.5}(a_i)$&  $93.6^{+5.1}_{-4.9}$\\
\hline \hline
\end{tabular}
\end{center}
\end{table}

\begin{table}[b]
\caption{Same as Table~\ref{tab:bra1-b1v} but for $B_c \to (K_1(1270)^+, K_1(1400)^+) (f_1(1285), f_1(1420))$ decays with
$\theta_3=38^\circ$(1st entry) and $\theta_3=50^\circ$(2nd entry).}
\label{tab:brk1a-k1bfa}
\begin{center}\vspace{-0.6cm}
\begin{tabular}[t]{l|lc|lc}
 \hline \hline
 $\Delta S =1$     &   \multicolumn{2}{|c|}{ $\theta_{K}=45^\circ$ }     &   \multicolumn{2}{|c}{ $\theta_{K}=-45^\circ$} \\
 Decay modes       &    BRs $(10^{-7})$  &  LPFs $(\%)$                &    BRs $(10^{-7})$  &  LPFs $(\%)$  \\
\hline
 $\rm{B_c \to {K_1}(1270)^{+} f_1(1285)}$                  &$1.4^{+0.9}_{-0.4}(m_c)^{+2.0}_{-0.7}(a_i)$&  $65.1^{+27.4}_{-19.4}$
                                             &$1.6^{+0.1}_{-0.5}(m_c)^{+1.1}_{-1.0}(a_i)$&  $96.7^{+2.7}_{-11.6}$\\
                                             &$1.7^{+1.1}_{-0.4}(m_c)^{+2.3}_{-1.0}(a_i)$&  $69.1^{+22.1}_{-19.6}$
                                                           &$1.5^{+0.3}_{-0.6}(m_c)^{+1.6}_{-1.2}(a_i)$&  $92.1^{+2.8}_{-13.0}$\\
 $\rm{B_c \to {K_1}(1400)^+ f_1(1285)}$                &$1.5^{+0.2}_{-0.4}(m_c)^{+1.2}_{-0.8}(a_i)$&  $96.7^{+2.7}_{-11.5}$
                                             &$1.4^{+0.8}_{-0.4}(m_c)^{+1.8}_{-0.8}(a_i)$&  $65.5^{+27.2}_{-19.4}$\\
                                             &$1.5^{+0.3}_{-0.6}(m_c)^{+1.6}_{-1.2}(a_i)$&  $92.1^{+4.0}_{-12.8}$
                                                       &$1.7^{+1.1}_{-0.5}(m_c)^{+2.2}_{-1.0}(a_i)$&  $69.5^{+21.9}_{-19.6}$\\
  \hline
 $\rm{B_c \to {K_1}(1270)^+ f_1(1420)}$                &$0.9^{+0.4}_{-0.3}(m_c)^{+0.8}_{-0.9}(a_i)$&  $81.6^{+13.5}_{-34.6}$
                                             &$4.4^{+0.6}_{-0.4}(m_c)^{+1.5}_{-1.7}(a_i)$&  $71.5^{+4.8}_{-8.9}$\\
                                             &$0.6^{+0.1}_{-0.2}(m_c)^{+0.4}_{-0.6}(a_i)$&  $78.5^{+16.9}_{-48.1}$
                                                       &$4.4^{+0.5}_{-0.3}(m_c)^{+1.2}_{-1.5}(a_i)$&  $73.2^{+4.8}_{-9.3}$\\
 $\rm{B_c \to {K_1}(1400)^+ f_1(1420)}$              &$4.3^{+0.6}_{-0.4}(m_c)^{+1.6}_{-1.7}(a_i)$&  $71.9^{+4.8}_{-9.3}$
                                             &$0.9^{+0.4}_{-0.3}(m_c)^{+0.8}_{-0.9}(a_i)$&  $81.9^{+13.2}_{-34.4}$\\
                                             &$4.4^{+0.5}_{-0.3}(m_c)^{+1.1}_{-1.6}(a_i)$&  $73.6^{+4.8}_{-8.8}$
                                                     &$0.6^{+0.1}_{-0.2}(m_c)^{+0.4}_{-0.7}(a_i)$&  $78.7^{+16.8}_{-47.3}$\\
\hline \hline
\end{tabular}
\end{center}
\end{table}

\begin{table}[b]
\caption{Same as Table~\ref{tab:bra1-b1v} but for $B_c \to (K_1(1270)^+, K_1(1400)^+) (h_1(1170), h_1(1380))$ decays with
$\theta_1=10^\circ$(1st entry) and $\theta_1=45^\circ$(2nd entry).}
\label{tab:brk1a-k1bha}
\begin{center}\vspace{-0.6cm}
\begin{tabular}[t]{l|lc|lc}
  \hline    \hline
 $\Delta S =1$     &   \multicolumn{2}{|c|}{ $\theta_{K}=45^\circ$ }     &   \multicolumn{2}{|c}{ $\theta_{K}=-45^\circ$} \\
 Decay modes       &    BRs $(10^{-6})$  &  LPFs $(\%)$                &    BRs $(10^{-6})$  &  LPFs $(\%)$  \\
 \hline
 $\rm{B_c \to {K_1}(1270)^+ h_1(1170)}$         &$1.4^{+0.6}_{-0.6}(m_c)^{+1.3}_{-0.8}(a_i)$&  $94.5^{+2.3}_{-3.9}$
                                  &$1.6^{+0.7}_{-0.5}(m_c)^{+1.0}_{-1.0}(a_i)$&  $98.5^{+0.6}_{-0.9}$\\
                                  &$0.6^{+0.3}_{-0.3}(m_c)^{+0.3}_{-0.4}(a_i)$&  $87.9^{+6.5}_{-14.6}$
                                  &$0.2^{+0.2}_{-0.0}(m_c)^{+0.3}_{-0.0}(a_i)$&  $92.9^{+7.5}_{-15.1}$\\
 $\rm{B_c \to {K_1}(1400)^+ h_1(1170)}$     &$1.6^{+0.7}_{-0.5}(m_c)^{+1.0}_{-1.1}(a_i)$&  $98.5^{+0.6}_{-0.9}$
                                  &$1.4^{+0.6}_{-0.6}(m_c)^{+1.2}_{-0.9}(a_i)$&  $94.6^{+2.3}_{-3.9}$\\
                                  &$0.2^{+0.2}_{-0.0}(m_c)^{+0.3}_{-0.0}(a_i)$&  $93.0^{+7.3}_{-14.8}$
                                  &$0.5^{+0.4}_{-0.3}(m_c)^{+0.6}_{-0.2}(a_i)$&  $88.1^{+6.4}_{-14.4}$\\
\hline
 $\rm{B_c \to {K_1}(1270)^+ h_1(1380)}$     &$0.9^{+0.3}_{-0.0}(m_c)^{+0.8}_{-0.3}(a_i)$&  $98.5^{+0.8}_{-1.4}$
                                  &$1.5^{+0.5}_{-0.4}(m_c)^{+0.9}_{-0.7}(a_i)$&  $89.6^{+2.9}_{-4.0}$\\
                                  &$1.8^{+0.5}_{-0.4}(m_c)^{+1.1}_{-0.9}(a_i)$&  $98.6^{+0.8}_{-0.7}$
                                  &$2.8^{+1.1}_{-0.8}(m_c)^{+1.7}_{-1.3}(a_i)$&  $94.3^{+1.7}_{-2.9}$\\
 $\rm{B_c \to {K_1}(1400)^+ h_1(1380)}$ &$1.5^{+0.4}_{-0.4}(m_c)^{+0.8}_{-0.7}(a_i)$&  $89.8^{+2.8}_{-3.9}$
                                  &$0.9^{+0.3}_{-0.1}(m_c)^{+0.8}_{-0.4}(a_i)$&  $98.5^{+0.9}_{-1.3}$\\
                                  &$2.8^{+1.1}_{-0.8}(m_c)^{+1.6}_{-1.3}(a_i)$&  $94.4^{+1.7}_{-2.7}$
                                  &$1.7^{+0.6}_{-0.3}(m_c)^{+1.4}_{-0.7}(a_i)$&  $98.6^{+0.9}_{-0.5}$\\
\hline \hline
\end{tabular}
\end{center}
\end{table}


Based on the numerical results as given in Tables~\ref{tab:bra1-b1v}-\ref{tab:brk1a-k1bha}, 
some remarks are in order:
\begin{itemize}

\item Among  the considered sixty two pure annihilation $B_c\to AV/VA,
AA$ decays, the pQCD predictions for the {\it CP}-averaged BRs of
those $\Delta S = 0$ processes are generally much larger
than those of $\Delta S =1$ channels (one of the two final state
mesons is a strange meson),
the main reason is the enhancement of
the large CKM factor $|V_{ud}/V_{us}|^2 \sim 19$ for those $\Delta
S = 0$ decays as expected in general. Maybe there exists no such large differences for
certain decays, which is just because the enhancement arising from the
CKM factor is partially cancelled by the difference between the
magnitude of individual decay amplitude.

\item
There is no {\it CP} violation for all these sixty two decays within the SM, since there
is only one kind of tree operator involved in the decay amplitude of
all considered $B_c$ decays, which can be seen directly from Eq.~(\ref{eq:amt}).

\item
For the ten $B_c \to (a_1, b_1) V$ decays, the pQCD predictions of the BRs and
LPFs for both
$\Delta S=0$ and $\Delta S=1$ processes are listed
in the Table~\ref{tab:bra1-b1v}.
As argued in Ref.~\cite{ekou09:ncbc}, the LHCb experiments
could observe the BRs of annihilation $B_c$ meson decays at the level of $10^{-6}$,
the decays $B_c \to a_1 \omega, b_1 \rho$
will thus be detected at LHC because they are just within its reach.
As for the polarization, all these ten decays are governed
by the longitudinal contributions. The LPFs are around 95\% within the theoretical
errors except for $B_c \to a_1^+\omega, a_1 K^*$ modes ($\sim$ 85\%),
and  for $B_c \to a_1 \rho$ channels, $f_L(B_c \to a_1 \rho) \sim 1$,
which will be tested in the LHCb experiments.

\item
Since the behavior of $b_1$ meson is contrary to that of $a_1$ meson, 
one can find that $Br(B_c \to b_1 \rho) > Br(B_c \to a_1 \rho) $ as given in Table~\ref{tab:bra1-b1v}
 and the ratio of the corresponding BRs for $B_c \to a_1 \rho$
and $B_c \to b_1 \rho$ is that
\beq
\frac{Br(B_c \to b_1^+ \rho^0)}{Br(B_c \to a_1^+ \rho^0)}&=& \frac{Br(B_c \to b_1^0 \rho^+)}{Br(B_c \to a_1^0 \rho^+)}
\approx 13.2\;,\label{eq:rabr}
\eeq
Similarly, for $B_c \to a_1 K^*, b_1 K^*$ decays, the BRs of the latter modes are larger than that of the former ones and
\beq
\frac{Br(B_c \to b_1^+ {K^*}^0)}{Br(B_c \to a_1^+ {K^*}^0)}&=& \frac{Br(B_c \to b_1^0 {K^*}^+)}{Br(B_c \to a_1^0 {K^*}^+)}
 \approx 5.5\;,\label{eq:rabk}
\eeq
The above two ratios exhibit the annihilation decay pattern consistent with those
as shown in Ref.~\cite{Cheng08:aa}.

\item
Analogous to $B_c \to \rho^+ \rho^0$ decay~\cite{xiao09:bcnd}, the
contributions from $\bar u u$ and $\bar d d$ components cancel each
other exactly and result in the zero BRs for $B_c \to a_1^+ a_1^0$ and $B_c \to b_1^+ b_1^0$.
 Any other nonzero data for these two channels may
indicate the effects of exotic new physics.
While for $B_c \to a_1^+ b_1^0$ and $B_c \to b_1^+ a_1^0$, as expected
from the analytic expressions,
Eqs.~(\ref{eq:a1pb10},\ref{eq:b1pa10}), due to the same component
of $ u\bar u -  d\bar d$ involved in both axial-vector $a_1^0$ and
$b_1^0$ mesons at the quark level, the pQCD predictions for the
BRs and LPFs as given in Table~\ref{tab:bra1-b1a} show the identical results as they should be,
\beq
Br(B_c \to a_1^+ b_1^0)&=& Br(B_c \to b_1^+ a_1^0) \approx 2.2 \times 10^{-5}\;, \non
f_L(B_c \to a_1^+ b_1^0)&=& f_L(B_c \to b_1^+ a_1^0) \approx 92\%\;.
\eeq
where the large BRs
($\sim 10^{-5}$) are within the reach of the LHCb experiments~\cite{ekou09:ncbc} and could be
detected at LHC.

\item
Since the $^3P_1$ meson behaves like the vector meson and $f_{a_1} \sim f_\rho $ 
from Eq.~(\ref{eq:dconst}),
the pQCD predictions of BRs exhibit the good consistency generally for $B_c \to a_1^+ \omega$ and
$B_c \to \rho^+ \omega$, $B_c \to a_1^+ {K^*}^0$ and $B_c \to \rho^+ {K^*}^0$,
 $B_c \to a_1^+ b_1^0(a_1^0 b_1^+)$ and $B_c \to \rho^+ b_1^0(\rho^0 b_1^+)$ decays, respectively,
within the theoretical errors as roughly estimated. As for the polarizations, which can well
manifest the helicity structure 
for the corresponding modes, the LPFs
present the different features from the decay rates except
for $B_c \to a_1^+ b_1^0(a_1^0 b_1^+)$ and $B_c \to \rho^+ b_1^0(\rho^0 b_1^+)$ decays.
From Table~\ref{tab:bra1-b1v}, the LPFs for $B_c \to a_1^+ \omega$ and $B_c \to a_1^+ {K^*}^0$ can be read straight forward as:
$f_L(B_c \to a_1^+ \omega)=(84.7^{+5.0}_{-4.4})\%$ and $f_L(B_c \to a_1^+ {K^*}^0)=(83.6^{+5.3}_{-7.5})\%$.  As given in
Ref.~\cite{xiao09:bcnd}, $f_L(B_c \to \rho^+ \omega)=(92.9^{+2.0}_{-0.1})\%$ and $f_L(B_c \to \rho^+ {K^*}^0)=(94.9^{+2.0}_{-1.4})\%$,
where the various errors as specified have been added in quadrature. 
The above results and discussions would be tested with high precision by the relevant experiments
operated at the ongoing LHC and forthcoming Super-B to identify
the helicity structure even decay mechanism in these considered channels.

\item
In Table~\ref{tab:brf1-f8v}, from the pQCD predictions of the BRs and
LPFs for $B_c \to \rho^+ (f_1(1285), f_1(1420))(\Delta S =0)$ and
$B_c \to {K^*}^+ (f_1(1285), f_1(1420))(\Delta S =1)$ decays, one can
observe that the BRs of $B_c \to \rho^+ f_1(1420), {K^*}^+ f_1(1285)$
are more sensitive than those of $B_c \to \rho^+ f_1
(1285), {K^*}^+ f_1(1420)$ to the mixing angle $\theta_3$,
\beq
\frac{Br(B_c \to \rho^+ f_1(1420))|_{\theta_3=50^\circ}
}{Br(B_c \to \rho^+ f_1(1420))|_{\theta_3=38^\circ}}&\approx& 5.5\;;\label{eq:rf1}\\
\frac{Br(B_c \to {K^*}^+ f_1(1285))|_{\theta_3=38^\circ}}{
Br(B_c \to {K^*}^+ f_1(1285))|_{\theta_3=50^\circ}}&\approx& 4.0\;;\label{eq:kf1}
\eeq
These two relations, Eqs.~(\ref{eq:rf1},\ref{eq:kf1}), can be understood as that
the interferences between $B_c \to \rho^+ f_1(B_c \to {K^*}^+ f_1)$
and $B_c \to \rho^+ f_8(B_c \to {K^*}^+ f_8)$ become highly constructive(destructive)
to $B_c \to \rho^+ f_1(1420)(B_c \to {K^*}^+ f_1(1285))$ with the mixing
angle $\theta_3$ changing from $38^\circ$ to $50^\circ$.
Moreover,
\beq
\frac{Br(B_c \to \rho^+ f_1(1285))}{Br(B_c \to \rho^+ f_1(1420))}&\approx&
\left\{ \begin{array}{ll}
52.5& {\rm for \ \ \theta_3=38^\circ } \\
8.6& {\rm for \ \ \theta_3=50^\circ} \\ \end{array} \right.\;;\label{eq:rrhof}\\
\frac{Br(B_c \to {K^*}^+ f_1(1420))}{Br(B_c \to {K^*}^+ f_1(1285))}&\approx&
\left\{ \begin{array}{ll}
6.9& {\rm for \ \ \theta_3=38^\circ } \\
30.0& {\rm for \ \ \theta_3=50^\circ} \\ \end{array} \right.\;;\label{eq:rkstf}
\eeq
From the decay amplitudes as given in
Eqs.~(\ref{eq:rhopf'},\ref{eq:rhopf''},\ref{eq:kstf1},\ref{eq:kstf2}),
the  above two relations can be understood as follows:
(a) for $B_c \to \rho^+ (f_1(1285), f_1(1420))$ decays, the mixing coefficients for the former decay are
$\cos\theta_3$ and $\sin\theta_3$, while that for the latter one are $-\sin\theta_3$ and $\cos\theta_3$.
For the common component $q\bar q$, it is found that the contributions from $B_c \to \rho^+ f_1$ and $B_c \to \rho^+ f_8$ interfere
constructively(destructively) for $B_c \to \rho^+ f_1(1285)(B_c \to \rho^+ f_1(1420))$; (b) for $B_c \to {K^*}^+ (f_1(1285), f_1(1420))$ channels,
the mixing parameters remain unchanged, however, a new part of contribution from $s\bar s$ component involved in both $f_1$ and $f_8$ with
different signs results in the construction(destruction) to $B_c \to {K^*}^+ f_1(1420)(B_c \to {K^*}^+ f_1(1285))$.
Additionally, the LPFs for these decays are stable to the mixing
angle and play the dominant role except for $f_L(B_c \to {K^*}^+
f_1(1285))$, whose value change from $61.0\%$ at $\theta_3=38^\circ$ to
$33.7\%$ at $\theta_3=50^\circ$, which will be confronted with the
relevant experiments in the future.

\item
The pQCD predictions for $B_c \to (\rho^+, {K^*}^+) h_1$ decays, as given in Table~\ref{tab:brh1-h8v}, can be explained in a
similar way as for $B_c \to (\rho^+, {K^*}^+) f_1$.

\item
The numerical pQCD results for $\Delta S=0$ $B_c \to (a_1^+,
b_1^+) (f_1(1285), f_1(1420))$ decays, as given in
Table~\ref{tab:brf1-f8a}, can be commented in order:
(a) the BRs of these modes depend weakly on the mixing
angle $\theta_3$ except for $B_c \to a_1^+ f_1(1420)$; (b)
these 4 considered decays are governed by the longitudinal
contributions for both $\theta_3=38^\circ$ and
$\theta_3=50^\circ$; (c) as
mentioned in the text above, $1^3P_1$ meson behaves close to the vector
meson, the phenomenology of $B_c \to a_1^+ (f_1(1285), f_1(1420))$
can therefore be understood as that of $B_c \to \rho^+ (f_1(1285),
f_1(1420))$; (d) the mixing factors $\cos\theta_3$ and
$\sin\theta_3$ make the interference between $B_c \to (a_1^+,
b_1^+) f_1$ and $B_c \to (a_1^+, b_1^+) f_8$ constructive
(destructive) to $B_c \to (a_1^+, b_1^+) f_1(1285) (B_c \to
(a_1^+, b_1^+) f_1(1420))$.

\item
From the predictions for $B_c \to (a_1^+, b_1^+) (h_1(1170),
h_1(1380))$ presented in Table~\ref{tab:brh1-h8a}, some discussions could be addressed as: 
(a) since the behavior of $1^1P_1$ meson is
different even contrary to that of $1^3P_1$ meson, a surprisingly
large branching ratio for $B_c \to b_1^+ h_1(1700) (\sim 10^{-4})$
with the constructive effects induced by the
interference between $B_c \to b_1^+ h_1$ and $B_c \to b_1^+ h_8$
is produced, which will be tested stringently by the forthcoming relevant
LHC experiments; (b) once the large BRs are verified
by the measurements, the mixing angle $\theta_1$ could be well
determined 
to improve
the precision of the perturbative calculations; (c) except for $B_c \to b_1^+ h_1(1170)$ decay,
the rest three channels are sensitive significantly to the mixing
angle $\theta_1$; 
(d) similar to $B_c \to (a_1^+, b_1^+)(f_1(1285), f_1(1420))$ decays,
the longitudinal components play the dominant role for these 4
channels.

\item
For the $\Delta S =0$ $B_c \to \ov{K^*}^0 K_1^+$ and $B_c \to \ov{K_1} {K^*}^+$ decays,
one can see from Table~\ref{tab:brk1a-k1bv} that the BRs are large in the range of $10^{-6} \sim 10^{-5}$, which can be detected at the ongoing LHC
and forthcoming Super B experiments.
Moreover, the corresponding ratios of the BRs for these considered channels
 \beq
\frac{Br(B_c \to \ov{K^*}^0 K_1(1270)^+)}{Br(B_c \to \ov{K^*}^0 K_1(1400)^+)}&=& \frac{Br(B_c \to \ov{K_1}(1270)^0 {K^*}^+)}{Br(B_c \to \ov{K_1}(1400)^0 {K^*}^+)}
 \approx 1.7 \;, 
\eeq
for $\theta_K=45^\circ$, while
\beq
\frac{Br(B_c \to \ov{K^*}^0 K_1(1270)^+)}{Br(B_c \to \ov{K^*}^0 K_1(1400)^+)}&=& \frac{Br(B_c \to \ov{K_1}(1270)^0 {K^*}^+)}{Br(B_c \to \ov{K_1}(1400)^0 {K^*}^+)}
 \approx \frac{1}{1.7} \;,
\eeq
for $\theta_K=-45^\circ$, which indicate that one could determine the size and sign of the mixing angle $\theta_K$ after enough $B_c$ events
become available at the LHC experiments and then improve
 the precision of the theoretical predictions.
In terms of polarization, the longitudinal contributions play the dominated role for both
$\theta_K=45^\circ$ and $\theta_K=-45^\circ$ in $B_c \to \ov{K_1} {K^*}^+$ modes. 
In the $B_c \to \ov{K^*}^0 (K_1(1270)^+, K_1(1400)^+)$ decays, the transverse components govern the former channel for $\theta_K=-45^\circ$
while dominate the latter one for $\theta_K=45^\circ$.
These results will be tested by the relevant measurements in the future.


\item
Form the numerical results for $B_c \to K_1^+ (\rho, \omega)$, the $\Delta S =1$ processes, as displayed in the Table~\ref{tab:brk1a-k1bv},
one can straightforwardly observe that
\beq
Br(B_c \to K_1(1270)^+ \omega) &\sim & Br(B_c \to K_1(1270)^+ \rho^0)\non
 &=& \frac{1}{2}Br(B_c \to K_1(1270)^0 \rho^+)\;;\label{eq:brk127}\\
Br(B_c \to K_1(1400)^+ \omega) &\sim & Br(B_c \to K_1(1400)^+ \rho^0)\non
&=& \frac{1}{2}Br(B_c \to K_1(1400)^0 \rho^+)\;;\label{eq:brk140}
\eeq
\beq
f_L(B_c \to K_1(1270)^+ \omega) &\sim& f_L(B_c \to K_1(1270)^+ \rho^0)\non
 &=& f_L(B_c \to K_1(1270)^0 \rho^+) \;;\label{eq:flk127}\\ 
f_L(B_c \to K_1(1400)^+ \omega) &\sim & f_L(B_c \to K_1(1400)^+ \rho^0)\non
& =& f_L(B_c \to K_1(1400)^0 \rho^+)\;. \label{eq:flk140}
\eeq
within errors for both $\theta_K=45^\circ$ and $\theta_K=-45^\circ$, where the longitudinal components
contribute to these considered decays dominantly. The pattern of these decays shown in Eqs.~(\ref{eq:brk127}-\ref{eq:flk140})
can be understood as follows: only the same component $u\bar u$ in both of $\rho^0$ and $\omega$ mesons
contributes to these physical observables, where the differences
 mainly arise from the different decay constants.
Furthermore, by comparison with $B_c \to \ov{K^*}^0 K_1^+$
and $B_c \to \ov{K_1} {K^*}^+$ decays, one can find
that the pQCD predictions of $B_c \to K_1^+ (\rho, \omega)$ show the weak dependance
 on the value of the mixing angle $\theta_K$, which will also be tested by the LHC measurements.

\item
For $B_c \to K_1^+ \phi$ decays, it is interesting to note that
the decay rates, as listed in Table VII, are close to each other
within the theoretical uncertainties,
however, the LPFs show us the dramatically different features: the former is dominated by the longitudinal
components($\sim 95\%$) while the latter governed by the transverse ones($\sim 30\%$) when $\theta_K=45^\circ$.
The main reason is that the interferences induced by
$B_c \to K_{1A}^+ \phi$ and $B_c \to K_{1B}^+ \phi$ are constructive(destructive) to
$B_c \to K_1(1400)^+ \phi(B_c \to K_1(1270)^+ \phi)$ in the two transverse polarizations, 
meanwhile, these interferences in the longitudinal polarization contribute to these considered two decays oppositely.
When $\theta_K= -45^\circ$, the situation is quite the contrary.
The decay mechanism and helicity structure for $B_c \to K_1^+ \phi$ decays will be tested by the LHCb and Super-B experiments.

\item
For the $\Delta S =0$ processes, $B_c \to \ov{K_1} K_1^+$ modes, as presented in Table~\ref{tab:brk1a-k1ba},
it is of interest to notice that the BRs for all these four considered decays are in the order of $10^{-5}$, which are within the reach of $B_c$
experiments at LHC greatly as discussed in Ref.~\cite{ekou09:ncbc}. These numerical results also present the strong dependance on the mixing angle $\theta_K$,
which will also be tested by the relevant experiments in the near future.
The longitudinal polarization fractions are around $95\% \sim 100\%$ within theoretical errors except for $B_c \to \ov{K_1}(1400)^0 K_1(1400)^+$
($\sim 73\%$) at $\theta_K=45^\circ$ or $B_c \to \ov{K_1}(1270)^0 K_1(1270)^+$ ($\sim 72\%$) at $\theta_K=-45^\circ$ and paly the dominant role.

\item
As mentioned in the text above, although the suppressed CKM factor $V_{us} \sim 0.22$
is involved in the decay amplitudes(see Eqs.~(\ref{eq:k127a1},\ref{eq:k127b1},\ref{eq:k140a1},\ref{eq:k140b1}))
for the $\Delta S =1$ $B_c \to K_1 (a_1, b_1)$ decays, the pQCD predictions of the BRs
for $B_c \to K_1 b_1$ are in the order
of $10^{-6}$ and larger than that for $B_c \to K_1 a_1$ because the
$^1P_1$ meson behaves differently even contrarily to
the $^3P_1$ meson, whose behavior is close to that of the vector meson.
For the polarization fractions, all of these eight channels are governed by the longitudinal contributions evidently.
From the numerical
results as given in Table~\ref{tab:brk1a-k1ba}, one can also
find that the pQCD predictions of $B_c \to K_1 a_1(B_c \to K_1 b_1)$ are much(less) sensitive to the mixing angle $\theta_K$.

\item
For the $\Delta S =1$ $B_c \to K_1^+ (f_1(1285), f_1(1420))$ decays, from the pQCD
predictions presented in Table~\ref{tab:brk1a-k1bfa}, one can see
that the contributions to the BRs for these four decays come from the
overlap of various parts of $B_c \to K_{1A}^+ f_1, K_{1B}^+ f_1,
K_{1A}^+ f_8$, and $K_{1B}^+ f_8$, which have been given in
Eqs.~(\ref{eq:k127f'}-\ref{eq:k140f''}). Combining with four mixing
parameters $\cos\theta_K$, $\sin\theta_K$, $\cos\theta_3$ and
$\sin\theta_3$, these interferences result in the equivalent BRs
for $B_c \to K_1(1270)^+ f_1(1285)$ and $B_c \to K_1(1400)^+
f_1(1285)$ decays, and suppressed one for $B_c \to K_1(1270)^+
f_1(1420)$ while enhanced one for $B_c \to K_1(1400)^+
f_1(1420)$.  Moreover, (a) the BRs for $B_c \to
K_1^+ f_1(1420)(B_c \to K_1^+ f_1(1285))$ 
depend strongly(weakly)
on $\theta_K$ for both $\theta_3=38^\circ$ and
$\theta_3=50^\circ$; (b) the BRs for $B_c \to
{K_1}(1270)^+ f_1$ are more sensitive than that
for $B_c \to {K_1}(1400)^+ f_1$ to $\theta_3$
when $\theta_K=45^\circ$, while the situation is quite the
contrary when $\theta_K=-45^\circ$; (c) the longitudinal
contributions play an important role in all these considered
channels.

\item
Based on Eq.~(\ref{eq:k1mixing}), apart from an overall sign, the physical
states $K_1(1270)$ and $K_1(1400)$ can go one into another
with changing the mixing angle $\theta_K$ from $45^\circ$ to $-45^\circ$ and vice versa,
\beq
|K_1(1270)\rangle_{\theta_K=45^\circ}&=& |K_1(1400)\rangle_{\theta_K=-45^\circ},\non
|K_1(1400)\rangle_{\theta_K=45^\circ}&=& -|K_1(1270)\rangle_{\theta_K=-45^\circ}.
\label{eq:k1mix}
\eeq
which further results in the decay amplitudes of $B_c \to K_1 (V,A)$
(Here, $A$ is a nonstrange axial-vector meson) as follows:
\beq
{\cal A} (B_c \to K_1(1270) (V,A))_{\theta_K=45^\circ} &=& {\cal A} (B_c \to K_1(1400) (V,A))_{\theta_K=-45^\circ}\;, \label{eq:scheme1}\\
{\cal A} (B_c \to K_1(1400) (V,A))_{\theta_K=45^\circ} &=& -{\cal A} (B_c \to K_1(1270) (V,A))_{\theta_K=-45^\circ}\;.\label{eq:scheme2}
\eeq
These two relations, i.e., Eqs.~(\ref{eq:scheme1}) and~(\ref{eq:scheme2}),
can be manifested by the analytic
formulas for $B_c \to K_1 (V,A)$ decays as shown in
Eqs.~(\ref{eq:kstk127}-\ref{eq:k140kst}),
(\ref{eq:k127rho}-\ref{eq:k140ome}), (\ref{eq:phik127}-\ref{eq:phik140})
and~(\ref{eq:k127a1}-\ref{eq:k140h''}).
The pQCD predictions  for these considered $B_c \to K_1 (V,A)$ decays as listed in
the second and third columns of
Tables~\ref{tab:brk1a-k1bv},~\ref{tab:brk1a-k1ba},
~\ref{tab:brk1a-k1bfa} and~\ref{tab:brk1a-k1bha} also display the phenomenologies induced by the same pattern. 

\item
For $B_c \to \ov{K_1} K_1^+$ decays, however, it is not the case as shown
in Eqs.~(\ref{eq:scheme1}) and~(\ref{eq:scheme2}).
According to the relation shown in Eq.~(\ref{eq:k1mix}),
there are some simple relations between the decay amplitudes
as given in Eqs.~(\ref{eq:k127k127}-\ref{eq:k140k140})
for $B_c \to \ov{K_1} K_1^+$ decays :
\beq
{\cal A} (B_c \to \ov{K_1}(1270) K_1(1270))_{\theta_K=45^\circ} &=& -{\cal A} (B_c \to
\ov{K_1}(1400) K_1(1400))_{\theta_K=-45^\circ}\;, \label{eq:scheme3}\\
{\cal A} (B_c \to \ov{K_1}(1400) K_1(1400))_{\theta_K=45^\circ} &=& -{\cal A} (B_c \to
\ov{K_1}(1270) K_1(1270))_{\theta_K=-45^\circ}\;, \label{eq:scheme4}\\
{\cal A} (B_c \to \ov{K_1}(1270) K_1(1400))_{\theta_K=45^\circ} &=& {\cal A} (B_c \to
\ov{K_1}(1400) K_1(1270))_{\theta_K=-45^\circ}\;, \label{eq:scheme5}\\
{\cal A} (B_c \to \ov{K_1}(1400) K_1(1270))_{\theta_K=45^\circ} &=& {\cal A} (B_c \to
\ov{K_1}(1270) K_1(1400))_{\theta_K=-45^\circ}\;\label{eq:scheme6}.
\eeq
Of course, the above four relations, i.e., Eqs.~(\ref{eq:scheme3}-\ref{eq:scheme6}),
can also be extracted from the pQCD predictions of BRs
for $B_c \to \ov{K_1} K_1^+$ decays as presented in Table~\ref{tab:brk1a-k1ba} apart
from an overall sign.

\item
At the first sight, it appears that the numerical results
for $B_c \to (f_1, h_1) (V, A)$(Here, $A$ is either a $^3P_1$ or $^1P_1$ nonstrange axial-vector meson)
decays are determined by the mixing angles $\theta_3$ and
$\theta_1$, respectively, however, based on
Ref.~\cite{Yang07:twist}, whose values will be eventually
determined from $\theta_K$ in $K_{1A}$-$K_{1B}$ mixing system.
Experimentally, it is thus very important to measure the channels precisely
involving $K_1(1270)$ and/or $K_1(1400)$ 
to determine both
of sign and size of the mixing angle $\theta_K$ and reduce the uncertainties of theoretical predictions greatly.

\item
The pQCD predictions for the {\it CP}-averaged branching ratios of
considered $B_c$ decays vary in the range of $10^{-5}$ to
$10^{-9}$. Since the LHC experiment can measure the $B_c$ decays
with a branching ratio at $10^{-6}$ level~\cite{ekou09:ncbc}, our
pQCD predictions for the branching ratios of $B_c \to a_1^+ \omega$, $b_1 \rho$, $\ov{K^*}^0
K_1^+$, $\ov{K_1^0} {K^*}^+$, $\rho^+ f_1(1285)$, $a_1^+ b_1^0$, $b_1^+ a_1^0$, $a_1^+
f_1(1285)$, $a_1^+ h_1(1170)$, $b_1^+ h_1$,
$\ov{K_1^0} K_1^+$, $b_1^+ K_1^0$ and $K_1^+ h_1$
decays could be tested in the ongoing LHC experiments.

\item
It is worth stressing that the theoretical predictions in
the pQCD approach still have large theoretical errors induced by
the still large uncertainties of many input parameters, e.g.
Gegenbauer moments $a_i$. Any progress in reducing the error of
input parameters, such as the Gegenbauer moments $a_i$ and the
charm quark mass $m_c$, will help us to improve the precision of
the pQCD predictions.

\end{itemize}


\section{Summary}\label{sec:sum}

In summary, we studied the sixty two charmless hadronic $B_c \to VA, AA$
decays by employing the pQCD factorization approach based on the
$k_T$ factorization theorem systematically. These considered decay channels can only
occur via the annihilation type diagrams in the SM and they will provide an important platform
for testing the magnitude and decay mechanism of the annihilation contributions and understanding
the helicity structure of these considered channels and the content of the
axial-vector mesons. Furthermore, these decay modes might also reveal the existence of exotic new physics
scenario or nonperturbative QCD effects.

The pQCD predictions for the {\it CP}-averaged branching ratios and longitudinal polarization fractions are
displayed in Tables~(\ref{tab:bra1-b1v}-\ref{tab:brk1a-k1bha}).
From our perturbative evaluations and phenomenological analysis, we
found the following results:
\begin{itemize}

\item
The pQCD predictions for the branching ratios vary in the
range of $10^{-5}$ to $10^{-9}$. There are many charmless $B_c \to VA, AA$
decays with sizable branching ratios: $B_c \to a_1^+ \omega$, $b_1 \rho$, $\ov{K^*}^0
(K_1(1270)^+, K_1(1400)^+)$, $(\ov{K_1}(1270)^0, \ov{K_1}(1400)^0)
{K^*}^+$, $\rho^+ f_1(1285)$, $a_1^+ b_1^0$, $b_1^+ a_1^0$, $a_1^+
f_1(1285)$, $a_1^+ h_1(1170)$, $b_1^+ (h_1(1170), h_1(1380))$,
$(\ov{K_1}(1270)^0, \ov{K_1}(1400)^0)(K_1(1270)^+, K_1(1400)^+)$,
$b_1^+ (K_1(1260)^0, K_1(1400)^0)$ and $(K_1(1270)^+,
K_1(1400)^+)(h_1(1170), h_1(1380))$, which are with a decay rate at $10^{-6}$ or
larger and could be measured at the LHC experiment.

\item
For $B_c \to VA, AA$ decays, the branching ratios of  $\Delta
S= 0$ processes are generally much larger than those of $\Delta S =1$
ones. Such differences are mainly induced by the CKM factors
involved: $V_{ud}\sim 1 $ for the former decays while $V_{us}\sim
0.22$ for the latter ones.

\item
In general, since the behavior for $^1P_1$ meson is much different from
that for $^3P_1$ meson, the branching ratios of the pure annihilation
$B_c \to A(^1P_1) (V, A(^1P_1))$ are larger than that of $B_c \to
A(^3P_1) (V, A(^3P_1))$, which can be confronted with the LHC and Super-B
experiments.

\item
The longitudinal contributions play a dominant role in the most of these considered
pure annihilation $B_c \to VA, AA$ decays, which will be tested by the ongoing LHC
 and forthcoming Super-B experiments in the near future.

\item
The pQCD predictions for several decays involving mixtures of $^3P_1$ and/or $^1P_1$ mesons are rather sensitive to
the values of the mixing angles, which will be tested by the relevant experiments in the future.

\item
Because only tree operators are involved, the {\it CP}-violating asymmetries for
these considered $B_c$ decays are absent naturally.

\item
The pQCD predictions still have large theoretical
uncertainties, mainly induced by the uncertainties of the Gegenbauer
moments $a_i$ in the meson distribution amplitudes. By reducing
these uncertainties dramatically, one can improve the precision
of the theoretical predictions effectively.

\item
We here calculated the branching ratios and polarization fractions of
the pure annihilation $B_c \to VA, AA$ decays by employing the
pQCD approach. We do not consider the possible long-distance
contributions, such as the rescattering effects, although they should be present, and they
may be large and affect the theoretical predictions. It is beyond
the scope of this work.

\end{itemize}

\begin{acknowledgments}

X.~Liu would like to thank You-Chang~Yang for reading the manuscript.
This work is supported by the National Natural Science Foundation of China
under Grant No.10975074, and No.10735080, by the Project on Graduate
Students' Education and Innovation of Jiangsu
Province, under Grant No. ${\rm CX09B_{-}297Z}$,  and by the Project on
Excellent Ph.D Thesis of Nanjing Normal University, under Grant
No. 181200000251.

\end{acknowledgments}



\end{document}